\documentclass[twocolumn]{aastex631}

\newcommand{\SatGen}{\texttt{SatGen}}

\newcolumntype{?}{!{\vrule width 1pt}}

\usepackage{booktabs, tabularx, makecell, multirow, xspace}
\newcolumntype{C}[1]{>{\centering\let\newline\\\arraybackslash\hspace{0pt}}m{#1}}

\usepackage{boldline} 

\newcommand{\roughly}{\ensuremath{ {\sim}\,} }

\DeclareUnicodeCharacter{2212}{-}

\shorttitle{}
\shortauthors{}

\graphicspath{{./}{figures/}}
\usepackage{url}
\usepackage[normalem]{ulem}
\usepackage{amsmath}
\usepackage{comment}
\usepackage{svg}
\usepackage{natbib}
\usepackage{fontawesome}
\usepackage{subfigure}

\begin{document}

\title{\vspace{-40pt} StreamGen: Connecting Populations of Streams and Shells to Their Host Galaxies}

\correspondingauthor{Adriana Dropulic}
\email{dropulic@princeton.edu}

\author[0000-0002-7352-6252]{Adriana Dropulic}
\affiliation{Department of Physics, Princeton University, Princeton, NJ 08544, USA }

\author[0000-0003-2497-091X]{Nora Shipp}
\affiliation{Department of Astronomy, University of Washington, Seattle, WA 98195, USA}

\author[0000-0001-7052-6647]{Stacy Kim}
\affiliation{Carnegie Observatories, 813 Santa Barbara Street, Pasadena, CA 91101, USA}

\author[0009-0003-6359-5603]{Zeineb Mezghanni}
\affiliation{Department of Physics, Grinnell College, Guildford, Grinnell, IA 50112, USA}

\author[0000-0003-2806-1414]{Lina Necib}
\affiliation{Department of Physics and Kavli Institute for Astrophysics and Space Research, MIT, Cambridge, MA 02139, USA}
\affiliation{The NSF AI Institute for Artificial Intelligence and Fundamental Interactions, Cambridge, MA 02139, USA}

\author[0000-0002-8495-8659]{Mariangela Lisanti}
\affiliation{Department of Physics, Princeton University, Princeton, NJ 08544, USA }
\affiliation{Center for Computational Astrophysics, Flatiron Institute, 162 Fifth Ave, New York, NY 10010, USA}

\begin{abstract}
In this work, we study how the abundance and dynamics of populations of disrupting satellite galaxies change systematically as a function of host galaxy properties. 
We apply a theoretical model of the phase-mixing process to classify intact satellite galaxies, stellar stream-like and shell-like debris in $\roughly 1500$ Milky Way-mass systems generated by a semi-analytic galaxy formation code, \SatGen.  
In particular, we test the effect of host galaxy halo mass, disk mass, ratio of disk scale height to length, and stellar feedback model on disrupting satellite populations. We find that the counts of tidal debris are consistent across all host galaxy models, within a given host mass range, and that all models can have stream-like debris on low-energy orbits, consistent with those observed around the Milky Way. However, we find a preference for stream-like debris on lower-energy orbits in models with a thicker (lower-density) host disk or on higher-energy orbits in models with a more-massive host disk. Importantly, we observe significant halo-to-halo variance across all models. These results highlight the importance of simulating and observing large samples of Milky Way-mass galaxies and accounting for variations in host properties when using disrupting satellites in studies of near-field cosmology. 
\end{abstract}

\section{Introduction}
\label{sec:intro}

In the $\Lambda$ Cold Dark Matter~($\Lambda$CDM) paradigm, galaxies form through the process of hierarchical structure formation~\citep{white_galaxy_1991}, accumulating mass though the accretion and disruption of other, smaller galaxies. Observational signatures of this process include largely intact satellite galaxies, like the Milky Way's LMC~\citep{Putman_1998}, as well as the luminous remnants of satellite galaxies tidally disrupted by the host's gravitational potential; for reviews, see~\cite{Helmi_2020,bonaca2024stellar}. This debris falls broadly into two morphological categories: stream-like debris, which are coherent structures of tidally stripped stars that roughly follow their progenitor's orbit, and shell-like debris, which are more phase-mixed structures that form concentric arcs of stars perpendicular to their progenitor's orbit. 

Recent observations of surviving satellite galaxy populations, in the Milky Way and in external galaxies, have provided critical insight into the properties of dark matter and the processes involved in small-scale galaxy formation; for a review, see~\cite{Bullock_2017}. These observations have produced some of the the strongest constraints on galaxy formation at small scales and the properties of dark matter, but also revealed questions, including the missing satellites~\citep{moore1999, klypin1999}, core-cusp~\citep{navarro1996}, too-big-to-fail~(TBTF)~\citep{boylankolchin2011,boylankolchin2012,boylankolchin2023}, and the plane-of-satellites~\citep{Pawlowski_2018,kroupa_2005} problems. A partial resolution to the remaining questions has been achieved by modifying cosmological simulations to include baryonic physics~\citep{wetzel2016,garrisonkimmel2017,fitts2017,simpson2018,buck2019,Kim_2018,garrisonkimmel2019a,sales2022baryonic}. Tidal stripping of satellites is a critical component of this resolution, as it reduces the number of satellites and their central densities~\citep{Zolotov_2012,Brooks_2013,Brooks_2014,garrisonkimmel2017,garrisonkimmel2019a}, yet it is a large source of uncertainty---numerically, theoretically, and observationally---in various studies~\citep{VDB_2018,VDB_O_2018,Carlsten_2020,Errani_Penarrubia_2020}. The study of tidal disruption is additionally important because theories of dark matter and galaxy evolution can be constrained by tidal disruption rates~\citep{penarrubia_2012,errani_2015,Errani_2022b}. This is because the tidal disruption of satellite galaxies is highly sensitive to their underlying density profiles, which are in turn influenced by the particle nature of dark matter~\citep{Du_2018, Tulin_2018,Kaplinghat_2020,Eckert_2022,Shen_2022} and baryonic physics~\citep{garrisonkimmel2019a}. It is therefore essential to gain a more complete understanding of tidal stripping and satellite disruption---throughout entire halos of galaxies and at all stages of tidal disruption---through observation, theoretical modeling, and simulation.

The study of debris structures has become possible with observational and theoretical advances in detecting and modelling stripped satellites. The recent discoveries of numerous streams in the Milky Way~\citep[e.g.][]{Shipp_2018, Li_streams, Malhan_2022, Mir_Carretero_2023} and several candidates in external galaxies~\citep[e.g.][]{martinezdelgado2009,crnojevic2016,morales2018,kado-fong2018} have opened the opportunity for studying the disruption rate of satellite galaxies. Further, there is evidence of shell-like structures in the Galaxy~\citep{Helmi_2003,Deason_2013,Donlon_2020} and in other galaxies~\citep[for example,][]{Martinez_Delgado_2010,duc_2014,2022A&A...660A..28B,Dey_2023}. However, until recently, there has not been a straightforward description of the conditions that cause tidal debris to form a shell rather than a stream. Massive, radial mergers have been shown to preferentially produce shell-like debris in simulations~\citep{Amorisco_2015,Pop_2018,10.1093/mnras/stz1251}. \cite{Amorisco_2015} derived key parameters and mechanisms that dictate the morphology of tidal stream-like debris, including the mass ratio of merging galaxies, orbital parameters, dynamical friction, accretion time, and host galaxy potential. Later, \cite{hendel_and_johnston} crucially connected the formation of stream-like debris to that of shell-like debris through a theoretical model of the phase-mixing process as a function of the merging satellites’s mass, orbit, and  interaction time, as well as the host galaxy's properties.

Significant advances in modeling tidal debris have also been made in cosmological hydrodynamical simulations~\citep[e.g.,][]{Hopkins_2015,Hopkins_2018,Hopkins_2022, 10.1093/mnras/stx071, agertz2021vintergatan,Applebaum_2021} that have provided unprecedented insight into stream formation~\citep[][Riley et al. 2024 in prep., Shipp et al. 2024 in prep.]{Panithanpaisal_2021,Arora_2022,Shipp_2023}. \cite{Shipp_2023} identified populations of detectable stream-like debris in the Feedback In Realistic Environments~(FIRE) simulations~\citep{Hopkins_2015,Hopkins_2018} and found a potential discrepancy between simulations and observations---the stream-like debris identified in 13 Milky Way analogs consistently have higher pericenters and apocenters than the Milky Way stream-like debris. This disagreement could be due to any property that affects the tidal disruption of satellites, for example the host galaxy's properties (such as its total mass and structure),  the implementation of baryonic feedback, the resolution of the simulations, or the dark matter physics. For example, it has been shown in the literature that host properties, such as disk and halo mass, can have significant impact on satellite survivability, diversity, and  orbits~\citep{garrisonkimmel2017, Jiang_2021}, and will thus necessarily impact the population of stream-like debris and shell-like debris. 

However, the cosmological hydrodynamical simulations that investigate these questions about satellite tidal disruption are computationally expensive, making it challenging to disentangle the effects of various factors on the tidal debris population, including subgrid modeling in the simulations, the assumed dark matter model, and halo-to-halo variance. These systematics may cause the differences between observations and simulations of the Milky Way, which could be misinterpreted as small-scale structure discrepancies with $\Lambda \mathrm{CDM}$. Sophisticated models that generate populations of stream-like and shell-like debris are crucial for fully understanding how host galaxy properties influence tidal disruption.

Semi-analytic tools are an efficient way to rapidly generate multiple realizations of disrupted satellite populations, varying host properties~\citep{koposov_quantitative_2009, li_nature_2010, maccio_luminosity_2010, guo_dwarf_2011,font_population_2011,Benson_2012,brooks_baryonic_2013,starkenburg_satellites_2013,barber_orbital_2014,pullen_nonlinear_2014,guo_milky_2015,lu_connection_2016,Jiang_2016,nadler_modeling_2019,Yang_2020,Jiang_2021,nadler_symphony_2023}. In this work, we extend \SatGen, a semi-analytical satellite galaxy generator that can efficiently produce satellite populations for a large number of host halos~\citep{Jiang_2021}. We evolve each satellite in the dynamic potential of the host and apply a theoretical model of phase mixing to each satellite's orbit to estimate whether it is intact or if its debris would be stream-like or shell-like.  These broad steps comprise a debris identification pipeline for satellite galaxies, which we call \texttt{StreamGen}. The pipeline is based on an analytic model by \citet{hendel_and_johnston} (hereafter HJ15), which describes how satellites phase mix over time. We apply the ``morphology metric" from HJ15 to the \SatGen\xspace satellite orbits to generate stream-like and shell-like tidal debris populations in $\roughly 1500$ Milky Way-mass galaxies. We vary the host's mass, disk mass, and disk height around current best estimates of Milky Way values. For a given mass range of host galaxy, we find that the variations in disk properties that we selected do not significantly affect the orbital properties of stream and shell populations, especially when compared to halo-to-halo variance. 

This paper is organized as follows. Section~\ref{sec:sims} introduces \texttt{StreamGen}, our debris morphology identification code, including a brief overview of the initial \SatGen\xspace satellite galaxy generator and our extension to it to predict which satellites turn into stream-like or shell-like debris based on our implementation of the HJ15 morphology metric. Section~\ref{sec:results} uses the results from \texttt{StreamGen} to demonstrate how stream-like and shell-like populations vary across Milky Way-mass galaxies with varying properties. In Section~\ref{sec:stream_props}, we present the fiducial model, and in Section~\ref{sec:vary_host_sim}, we explore the effect of host halo and disk properties, as well as stellar feedback models, on stream-like and shell-like populations. We conclude in Sec.~\ref{sec:conclusion} with a discussion comparing our results to stream populations in cosmological simulations and a discussion of future work in light of upcoming surveys. Appendix~\ref{app:uncertainty} describes the effect of uncertainty in the morphology metric assignment on our results. Appendix~\ref{app:additional_plots} contains additional plots of potential interest to the reader.

Our debris identification code, \texttt{StreamGen}, which is built to classify tidal debris populations in \texttt{SatGen}, but in principle could be applied to the outputs from any other satellite generator, is available on \href{https://github.com/adropulic/StreamGen.git}{\texttt{GitHub}}. 

\section{Simulations} \label{sec:sims}
In this section, we detail the steps required to generate populations of tidal debris structures from a semi-analytical satellite generator. Section~\ref{sec:SatGen} reviews the generation of merger trees as well as the choice of a subhalo's initial conditions and its subsequent orbital evolution, as modeled by \SatGen. Section~\ref{morph_metric} discusses the definition of a morphology metric as the basis for identifying stream-like and shell-like debris given a satellite's properties. Section~\ref{sec:streamgen} presents the implementation of this model for \SatGen, which we call \texttt{StreamGen}. Section~\ref{sec:validate_metric} validates \texttt{StreamGen} using N-body simulations.

\subsection{SatGen}\label{sec:SatGen}
~\cite{Jiang_2021} presents \SatGen, a semi-analytic model of subhalo evolution within a host halo. The semi-analytic nature of the model permits the efficient generation of a large statistical sample of galaxies and their subhalos. A goal of this work is to classify what these subhalos look like at redshift $z=0$: whether they remain largely intact or if they have tidal debris that is predominantly stream-like or shell-like. \SatGen\xspace has a variety of tuneable properties that can emulate different cosmological hydrodynamical simulations, such as the structure of the host galaxy and the strength of stellar feedback, which can affect tidal disruption of satellites. These properties and the selections used in this work are described in this section. 

\SatGen\xspace creates satellite populations in a two-step process: it generates halo merger trees and subsequently evolves satellite orbit and structure. In \SatGen, the halo merger trees are generated using an algorithm based on the extended Press-Schechter~(EPS) formalism~\citep{1993MNRAS.262..627L}. Given a target halo of known mass and redshift, EPS gives the expected number and mass of progenitor halos at any earlier redshift.  Using the EPS formalism allows for the rapid random sampling of assembly histories, which is useful for studying halo-to-halo variance~\citep{Jiang_2021}. The host mass is given by the merger tree branch that tracks the most massive progenitor. The mass of the disk is a user-specified fraction of the host mass;~\cite{Jiang_2021} note that these disks mimic, but do not reproduce, the Milky Way's disk and are similar to the disks in the simulations of~\cite{Penarrubia_2010}.

After the merger trees are defined, the orbits and baryonic properties of satellites are initialized. Orbits of satellite galaxies are modeled based on distributions motivated by cosmological simulations ~\citep{Li_2020}, which  find that infall velocities of satellites  follow a nearly-universal distribution peaking near the host's virial velocity.\footnote{We note that the results of this work could be sensitive to the infall velocity distribution, which is in turn dependent on the cosmology, but we did not explicitly examine this in this work.} Properties such as the stellar mass and stellar size are assigned to subhalos at infall using the stellar-to-halo mass relation~(SMHM), assuming a scatter of 0.15~dex in $M_*$ at a given $M_{\mathrm{vir}}$~\citep{Rodr_guez_Puebla_2017}. We note that observations of nearby satellite systems indicate this may significantly underestimate the scatter in the SMHM relation in subhalos with $M_\mathrm{vir} \lesssim 10^{11} M_\odot$ \citep[e.g.][and references therein]{Danieli_2023}, which would affect subhalo evolution.  The stellar half-mass radius is evolved without assuming an underlying density profile for the stars~\citep{10.1093/mnras/stz1499}.

\SatGen~can emulate the baryonic feedback of the Numerical Investigation of Hundred Astrophysical Objects~(NIHAO) simulations~\citep{10.1093/mnras/stv2856,10.1093/mnras/staa2790}, which is characterized by repeated, short bursts of star formation.  It can also emulate the feedback of the A Project Of Simulating The Local Environment~(APOSTLE) simulations~\citep{Sawala_2016}, which has a smoother progression of star formation. The ``bursty'' feedback of the NIHAO simulations can simultaneously make dwarf galaxies more puffy and centrally cored~\citep{El_Badry_2016}, causing them to be more susceptible to tidal disruption. The ``smooth'' feedback of the APOSTLE simulations does not significantly core dwarf galaxies and therefore leaves them more resistant to tidal disruption. \SatGen's halo response to star formation and feedback, which affects the tidal disruption of satellites, is  implemented through a numerical correction to the subhalo's concentration and its inner density slope before satellite infall, as derived from the respective hydrodynamical simulation; this is described in Sec.~\ref{sec:apostle}. 

In this work, all \texttt{SatGen} halos have a Dekel-Zhao profile~\citep{10.1093/mnras/stx486, 10.1093/mnras/staa2790}. The Dekel-Zhao density profile is presented as a function of  ${x \equiv r / r_{\mathrm{s}}}$, where $r$ is the galactocentric radius and ${r_{\mathrm{s}}=r_{\mathrm{vir}} / c}$ is the scale radius, which is  defined in terms of the concentration $c$ and virial radius $r_{\text {vir}}$. The Dekel-Zhao density profile is given by
\begin{equation}\label{densityprof}
\rho(x)=\frac{\rho_0}{x^\alpha\left(1+x^{1 / \beta}\right)^{\beta(\gamma-\alpha)}} \, ,
\end{equation}
where $\beta=2$ is chosen in \texttt{SatGen}, ${\gamma=3+\beta^{-1}=3.5}$,  $ {\alpha=-\mathrm{dln} \rho /\left.\mathrm{dln} r\right|_{r \rightarrow 0}}$ is the (negative of the) logarithmic density slope in the halo center. The numerator is given by ${\rho_0=\left[c^3(1-\alpha/3)/ f(c, \alpha)\right] \Delta_c \varrho_{\mathrm{crit}}}$ with ${f(x, \alpha)=\chi(x)^{2(3-\alpha)}}$ and ${\chi(x) \equiv x^{1 / 2} /\left(1+x^{1 / 2}\right)}$. $\varrho_{\mathrm{crit}}$ is the critical density and $\Delta_c$ is the spherical overdensity with respect to the critical density.
Additionally, galactic disks are described by the profile in~\cite{1975PASJ...27..533M}, hereafter the MN profile. An MN disk is defined using three parameters: the disk mass, $M_d$, a scale radius, $a$, and a scale height, $b$. Its density profile is given by 

\begin{equation}
\rho(R, z)=\frac{M_{\mathrm{d}} b^2}{4 \pi} \frac{a R^2+(a+3 \zeta)(a+\zeta)^2}{\zeta^3\left[R^2+(a+\zeta)^2\right]^{5 / 2}},
\end{equation}
where $\zeta = \sqrt{z^2 + b^2}$ and $R$, $\phi$, and $z$ are the cylindrical coordinates. 

\texttt{SatGen} evolves a satellite's orbit as well as its structure within the host, including dynamical friction~\citep{1943ApJ....97..255C}, tidal effects~\citep{Green_2021_tides}, and density-profile evolution~\citep[via a transfer function;][]{Green_2019}. In particular, two competing effects---tidal stripping and tidal heating---are dominant contributions to the evolution of satellite mass and structure. While tidal stripping reduces the size of a satellite by removing mass, tidal heating causes the satellite to expand by increasing its internal kinetic energy. The total mass loss to tides is given by 
\begin{equation}\label{eq:mass_loss}
    dm = \alpha_{se} [m - m(r_{\mathrm{tide}})] \frac{dt}{t_\mathrm{dyn}} \, ,
\end{equation}
where $\alpha_{se}$ is the stripping efficiency parameter, $m$ is the satellite's peak mass, $r_{\mathrm{tide}}$ is the satellite tidal radius, $dt$ is a timestep, and $t_{\mathrm{dyn}}$ is the host dynamical time at the satellite position. The host dynamical time is given by 
\begin{equation}\label{eq:dyn_time}
    t_\mathrm{dyn}(r) = \sqrt{\frac{3\pi}{16 G \Bar{\rho}(r)}} \, ,
\end{equation}
where $G$ is the universal gravitational constant and $\Bar{\rho}(r)$ is the average density of the host at the satellite position. 

The analytic description of a satellite's net structural change is complicated because the effect of tidal heating must be considered in tandem with that of tidal stripping. To this end, tidal-evolution tracks, obtained from numerical simulations, are used to  evolve subhalo density profiles as well as stellar masses and half-mass radii~\citep{Penarrubia_2010, Errani_2018,Errani_2021}. Qualitatively, a satellite's size initially increases with subhalo mass loss due to tidal heating and re-virialization and then decreases as stripping becomes more dominant~\citep{Jiang_2021}. The tidal tracks were verified by~\cite{Errani_2021} to be consistent with high-resolution N-body simulations down to $M(r<r_{\mathrm{mx}})/M_\mathrm{acc}(r<r_{\mathrm{mx}}) \approx 1/300$, where $r_{\mathrm{mx}} \approx 2.16 r_s$, $ r_s$ is the scale radius of the halo, and $M_\mathrm{acc}$ indicates the subhalo mass at accretion. 
 
The workflow described in this section can produce populations of satellites and their disruption rates around a range of Milky Way-like hosts. It is important to note a few limitations of the current \SatGen\xspace model, including that it assumes spherical symmetry for the host and satellites and does not model a few processes: the disk growth~(except as a function of the total virial mass of the host) or dynamics, the back reaction of the satellites on the host halo, and the interaction of satellites with each other. These processes will complicate the tidal disruption process and present an avenue for future development. 

\subsection{Morphology Metric}
\label{morph_metric}

HJ15 developed a method to classify tidal debris based on their orbital properties. HJ15 introduces a ``morphology metric" that quantifies the conditions required for stream-like and shell-like structures to form in terms of the satellite's mass, orbit, and interaction time, leveraging scaling relations that govern orbital debris evolution and a simple model of how orbits phase mix over time. This allows us to approximate whether tidal debris is stream-like or shell-like. HJ15 recognized that debris structures span a spectrum of morphologies from streams to shells, including a few structures that possess qualities of both (i.e. their ``umbrella" systems) or are difficult to classify by eye.  As such, we use the terminology ``stream-like" and ``shell-like" to denote this uncertainty. This section will describe the morphology metric as defined in HJ15; the implementation for \SatGen\xspace will be described in Sec.~\ref{sec:streamgen}. 

\begin{figure}[t]
    \centering
    \includegraphics[width=0.7\columnwidth]{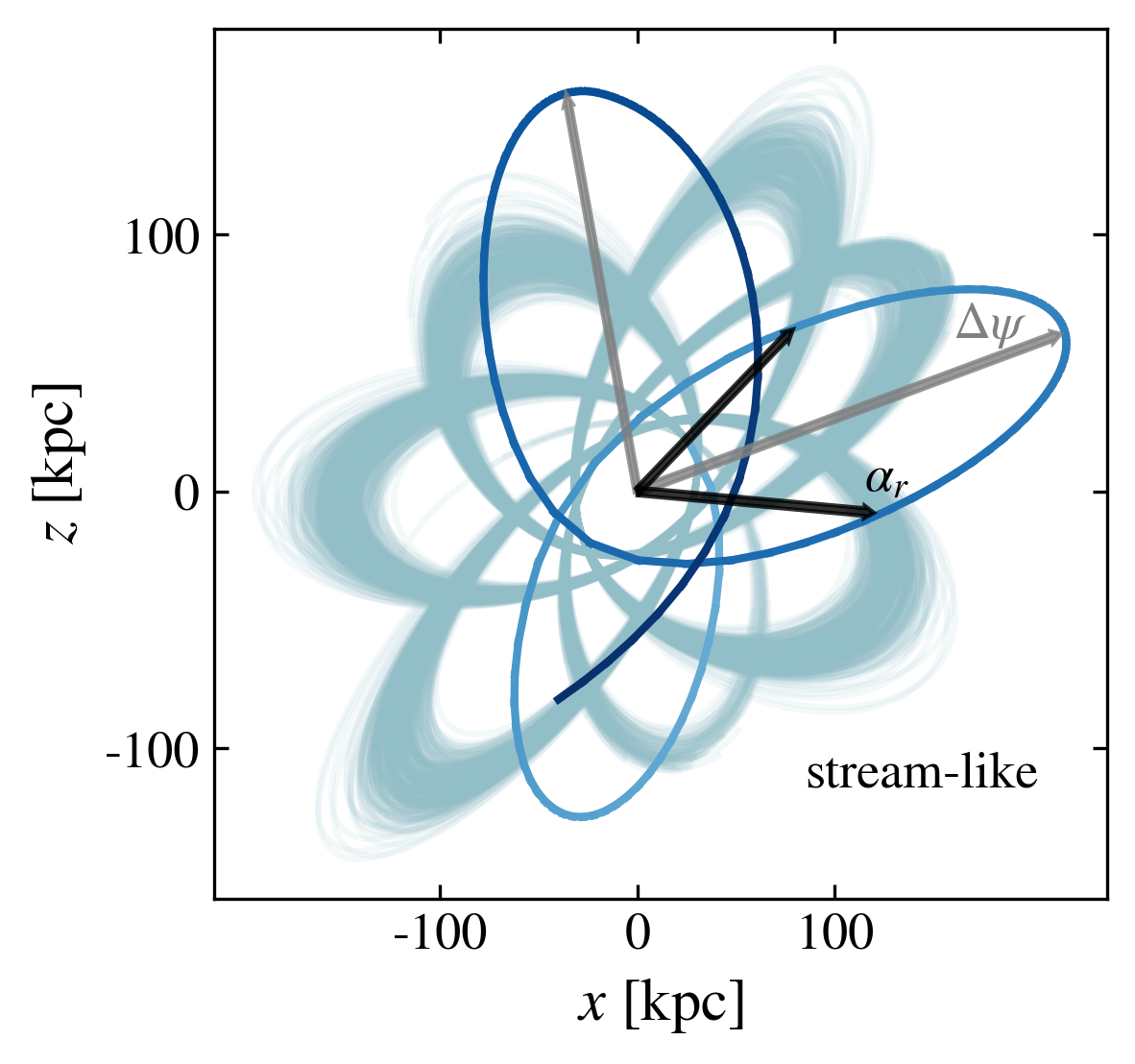} 
    \includegraphics[width=0.7\columnwidth]{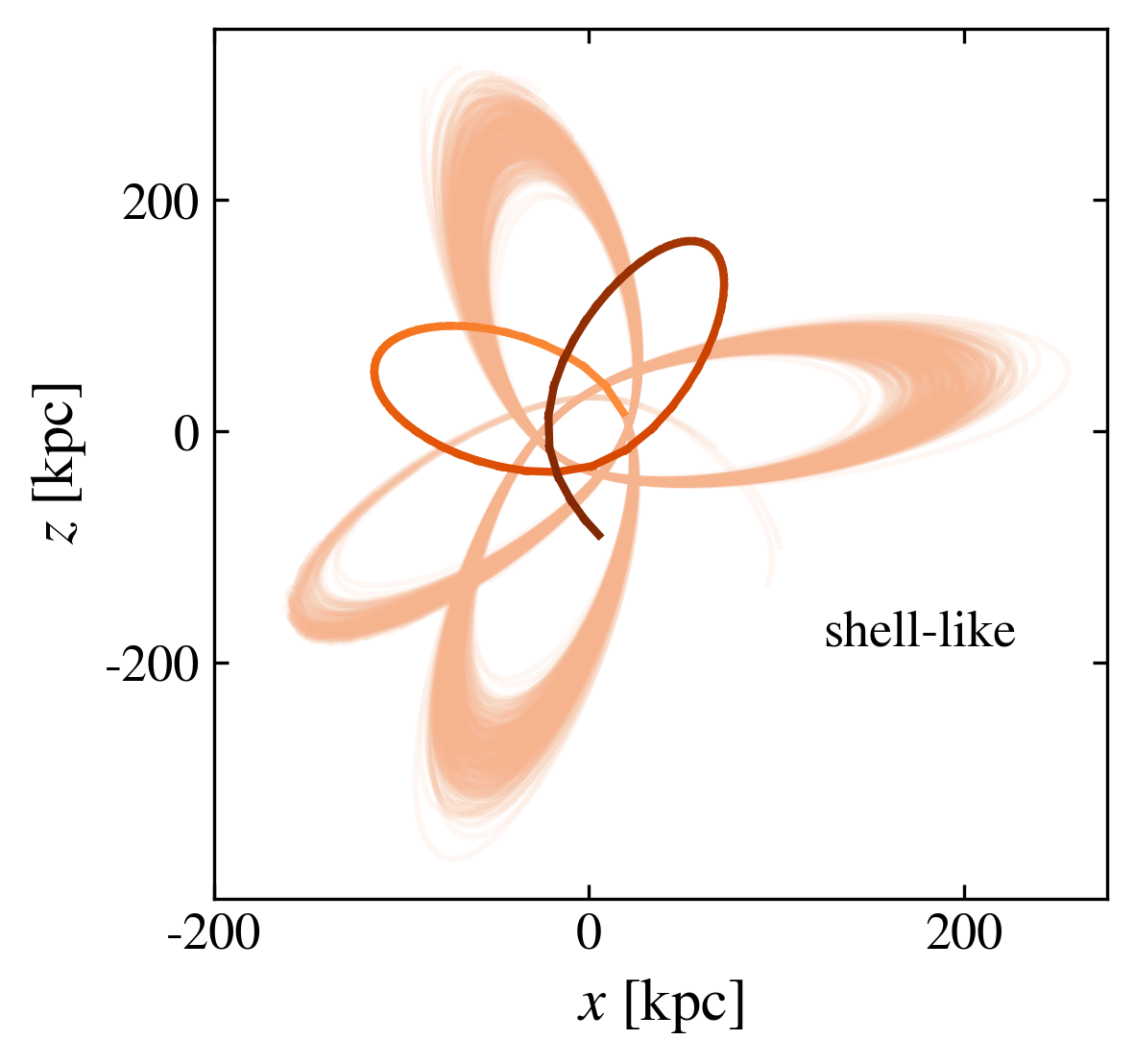}
    \caption{Orbits of two satellite galaxies in \texttt{SatGen} that are identified by the morphology metric as  stream-like~(blue) and shell-like~(orange). The dark, solid line is the original \SatGen\xspace orbit from satellite accretion until the point at which the morphology metric is calculated, which is 8.6~(4.8) Gyr for the stream~(shell). The light-to-dark gradient of this line indicates the time progression of the orbit from accretion until the point where the satellite has lost all but $1/300$ of its total initial mass. The thin lines in the background are the orbits of the ``stars” sampled from the \texttt{SatGen} satellite’s velocity dispersion and forward integrated from the point at which the morphology is calculated in the static host potential. The labels in the top panel correspond to the angle between apocenters~($\Delta \psi$) in grey and the rosette petal angle~($\alpha_r$) in black.}
    \label{fig:orbits}
\end{figure}

A variety of factors determine the shape and properties of a satellite galaxy's orbit: the total orbital energy~($E_{\rm orb}$), total orbital angular momentum~($L_{\rm orb}$), the host potential, the radii of the orbit at apocenter~($r_\mathrm{apo}$) and pericenter~($r_\mathrm{peri}$), the radial orbital period~($T_r$), the precession per orbit between successive apocenters~($\Delta \psi$) (referred to simply as the ``precession angle"), and the angular ``width" of a single rosette petal of the orbit~($\alpha_r$).  Alternatively, $\alpha_r$ is the angle through which the particle moves during the outer half of its radial period. Figure~\ref{fig:orbits} shows the orbits of two disrupting \SatGen\xspace satellite galaxies, in blue and orange, which will be discussed in detail later in this section. For illustrative purposes at this point, the angles $\Delta \psi$ and $\alpha_r$ are shown in projection for the \SatGen\xspace satellite orbit in the top panel of Fig.~\ref{fig:orbits}.

The energy~($E$) and angular momentum~($L$) dispersions for tidal debris are typically set at a satellite's pericenter, where satellite usually loses the most mass. In the HJ15 model, particles become unbound from the satellite through points approximately located at the satellite's (satellite-centric) tidal radius~($r_{\mathrm{tide}}$):
\begin{equation}\label{eq:tidal_radius}
    r_{\mathrm{tide}} = \left( \frac{m}{3M(r_\mathrm{peri})}\right)^{1/3} r_\mathrm{peri} \, ,
\end{equation}
where $m$ is the satellite's total dark matter mass at infall, and $M(r_\mathrm{peri})$ is the enclosed host mass at $r_\mathrm{peri}$. 

The difference in the host galaxy's gravitational potential energy across the satellite at pericenter determines the energy scale of the debris,

\begin{equation}\label{debris_energy_scale}
    e_s = 2 r_{\mathrm{tide}} \frac{\partial \Phi}{\partial r}\bigg|_{r_\mathrm{peri}} \, ,
\end{equation}
where $\Phi$ is the host potential.

The angular momentum scale is given by 
\begin{equation}\label{ang_mom_scale}
    l_s = (\sqrt{3} + 2) \left(\frac{m}{3M(r_\mathrm{peri})} \right)^{1/3}L_{\rm orb}.
\end{equation}

The angular spread of debris due to differences in azimuthal precession over time, $\Psi_L$, is given by 
\begin{equation}\label{psil}
\Psi_L = l_s N_{\rm orb} \frac{\partial\Delta\psi}{\partial L} \bigg|_{E_{\rm orb}},
\end{equation}
where $N_{\rm orb}$ is the number of orbits.

Stream-like variations in orbital period generate the angle $\Psi_E$, which is given by 
\begin{equation}\label{psie}
\Psi_E = e_s N_{\rm orb} \frac{\Delta\psi}{T_r}\frac{\partial T_r}{\partial E}\bigg|_{L_{\rm orb}}.
\end{equation}
$\Psi_E$ is restricted to be less than $\alpha_r$ so that only individual rosette petals are considered. The morphology metric is therefore defined as
\begin{equation}\label{mu}
\mu \equiv \bigg|\frac{\Psi_L}{\Psi^1_E}\bigg|    \, ,
\end{equation}
where 
\begin{equation}\label{min}
\Psi^1_E \equiv \text{min}(\alpha_r, \Psi_E).
\end{equation}

For dynamically-young debris, $\Psi^1_E > \Psi_L$. Eventually, $\Psi^1_E$ would become greater than $\alpha_r$, so when $\Psi^1_E = \alpha_r$, $\Psi^1_E$ remains constant at $\alpha_r$ thereafter. As the precession angle increases, angular momentum effects have a greater impact on the morphology. This continues until $\Psi_L = \Psi^1_E$, at which point the merger results in the formation of a shell-like structure. A faster transition occurs for higher-mass satellites. 

\subsection{StreamGen}\label{sec:streamgen}
In this section, we present our implementation of the morphology metric in \SatGen\xspace, which we call \texttt{StreamGen}. We explain how all inputs into the morphology metric, discussed in Sec.~\ref{morph_metric}, are calculated and examine the orbits, mass loss, and corresponding morphology metric values of two \SatGen\xspace satellite galaxies. Importantly, we consider satellites that have $M_* > 5 \times 10^5~M_\odot$ at accretion and do not fall within 4~kpc of the center of the primary host.\footnote{We do this to eliminate satellites whose orbits are susceptible to numerical effects at small radii $\lesssim 4$~kpc, as we find that as the satellite-host distance becomes small, the $1/r^2$ force can artificially kick the satellite out of the host. The radial resolution is 0.01~kpc, after which the satellite is considered to have merged fully into the host. } 

If a satellite has lost less than $10\%$ of its total stellar mass by $z=0$, it is considered to be intact. For satellites which have lost more mass and are considered disrupted, we calculate the morphology metric at satellite pericenter, as this is where most debris stripping occurs. We select the pericenter just preceding the point when the satellite has lost all but 1/300 of the dark matter mass within its virial radius,\footnote{We choose this point as an approximation to the limit to which \cite{Errani_2021}'s tidal tracks have been validated with simulations, which is down to 1/300 of the initial (pre-infall) mass within $r_{\rm mx}$. As mass loss is more severe in the outskirts, our cut on 1/300 of the mass left within the \emph{virial radius} rather than $r_{\rm mx}$ is conservative.} at which point a significant fraction of the stars will have been stripped; hereafter, we refer to this point as ``p300." We assume that the orbit of the remnant satellite at p300 well approximates the morphology of the \emph{stellar} debris. If the satellite does not reach p300, the morphology metric assignment occurs at the most recent pericenter. The time between p300 and the most recent pericenter can vary depending on multiple factors such as satellite mass, initial merger configuration, and accretion time.  We estimate the morphology of the stellar debris by applying the morphology metric at p300. In general, at this point, satellites can still retain $\sim 50\%$ of their stellar mass, which means that we are likely probing the morphology of this debris as it is stripped. This point is also significant because after the satellite has lost a greater fraction of its mass than it has at p300, the tidal tracks have not been explicitly validated against N-body simulations, although they still reproduce results from N-body simulations well~\citep{Errani_2021}. 

Next, we discuss the inputs to the morphology metric that are calculated at p300 or most recent pericenter, according to the paragraph above. To calculate the derivative quantities in $\Psi_L$ and $\Psi_E$ in the morphology metric, 300--1000 particles are sampled\footnote{We sample 1000 times if a satellite's $r_\mathrm{peri}< 20$~kpc and $r_\mathrm{apo}< 50$~kpc, suggesting that it is more likely to be disrupted. The sampling rate is set by requiring that the calculated derivative quantities are stable. } from the \SatGen\xspace satellite's velocity dispersion, $\sigma$:
\begin{equation}\label{sigma}
\sigma^2(x) = \frac{2 V_\mathrm{vir}^2 c}{f(c,\alpha)} \frac{x^{3.5}}{\chi^{2(3.5-\alpha)}} \sum_{i=0}^{i=8} \frac{(-1)^i 8! (1-\chi^{4(1-\alpha)+i})}{i! (8-i)! (4(1-\alpha)+i)} \,,
\end{equation}
where $V_\mathrm{vir}$ is the virial velocity, and the other quantities are as defined for Eq.~\ref{densityprof}. Eq.~\ref{sigma} is derived from~\cite{zhao_analytical_1996} (which follows~\cite{tremaine_family_1994}).
As this is $\sigma$ for a 1D velocity dispersion, assuming spherical symmetry and time independence, we take $\sigma$ to be the velocity dispersion in each Cartesian direction.\footnote{We do not consider anisotropy of the velocity dispersion, as we only utilize the sampling step to get an estimate of the derivative of orbital quantities, in Eq.~\ref{psil} and Eq.~\ref{psie}, calculated from the sampled particles.} Each particle is then drawn from the multivariate Gaussian distribution where the dispersion is given by Eq.~\ref{sigma} and the mean is given by the satellite's average velocity in \SatGen. 

To obtain orbits and orbital quantities of the sampled particles for the morphology assignment procedure, we evolve the particles for 2--3 orbital periods in the host potential at the time of p300. The total orbital energy is calculated as $E_{\rm orb} = \Phi(r_\mathrm{peri}) + \frac{1}{2} v_p^2$, and the orbital angular momentum is  $L_{\rm orb} = |\bf{r}_\mathrm{peri} \times \bf{v}_p|$ at first pericenter, where $\bf{v}_p$ is the satellite velocity at pericenter. The precession per orbit, $\Delta\psi$, between successive apocenters, $\bf{r}_{\rm apo,1}$ and $\bf{r}_{\rm apo,2}$, is calculated as 
\begin{equation}
\Delta\psi = 2\pi - \cos^{-1}\left(\frac{\bf{r}_{\rm apo,1} \cdot \bf{r}_{\rm apo,2}}{|\bf{r}_{\rm apo,1}||\bf{r}_{\rm apo,2}|}\right) \, ,
\end{equation} 
where the $2 \pi$ is necessary for the metric to be a positive value as in HJ15. $\Delta \Psi$ is illustrated in Fig.~\ref{fig:orbits}. Finally, the radial orbital period, $T_r$, is calculated as the time  between consecutive apocenters, ${t_{\bf{r}_{\rm apo,1}} - t_{\bf{r}_{\rm apo,2}}}$. To calculate the derivative quantities in the morphology metric, in Eq.~\ref{psil} and Eq.~\ref{psie}, we evaluate the slope of $\Delta\psi$ versus $L$ of particles at the $E_{\rm orb}$ of the original \SatGen\xspace satellite. Similarly, we evaluate the slope of $T_r$ versus $E$ of particles at the $L_{\rm orb}$ of the original \SatGen\xspace satellite. These values are input into the morphology metric described in Sec.~\ref{morph_metric} to classify \SatGen\xspace debris as stream-like or shell-like. 

\begin{figure}[t]
    \centering
    \includegraphics[width=\columnwidth]{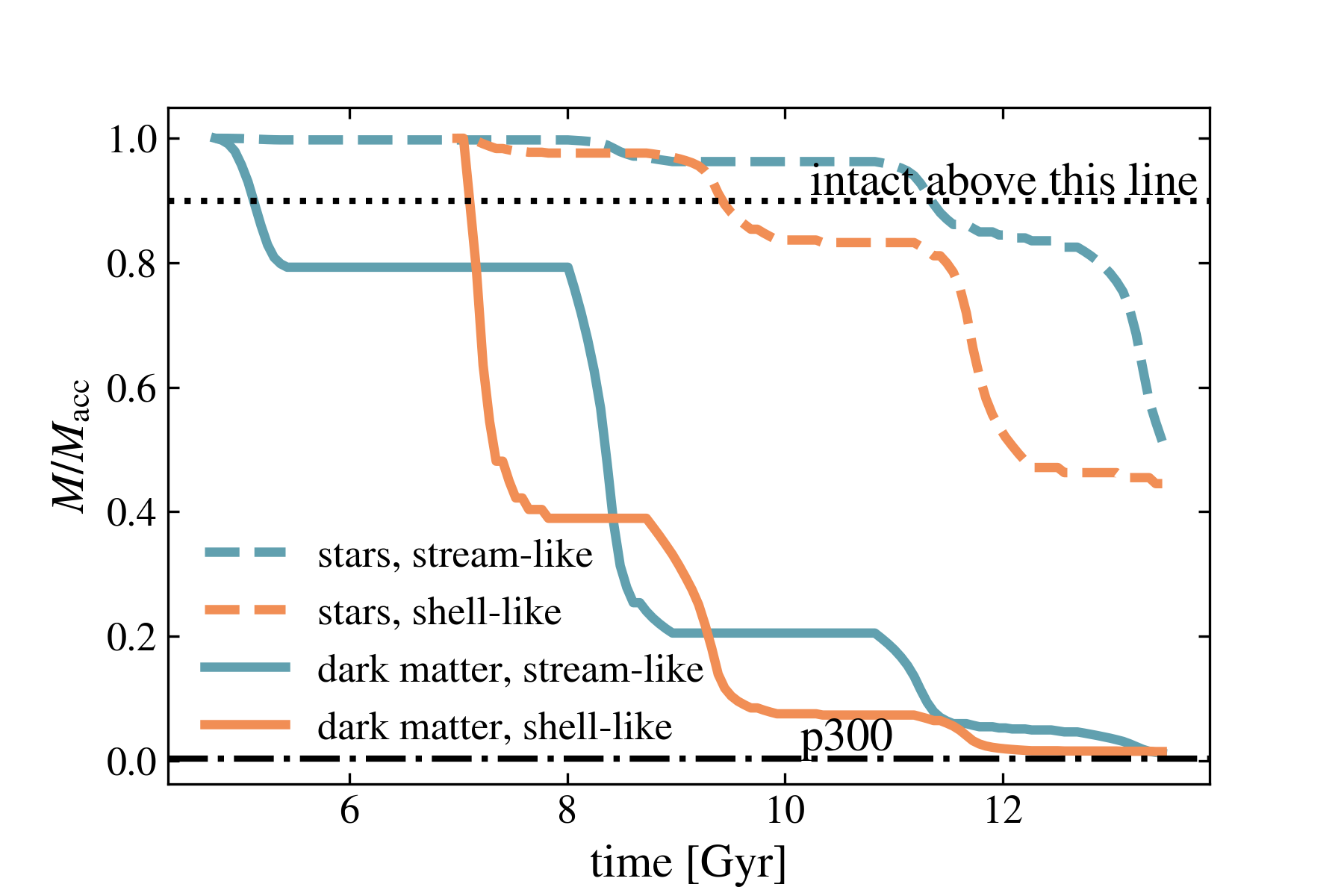}
\caption{Mass loss for dark matter~(solid lines) and stars~(dashed lines) over time for the satellites with stream-like debris~(blue) and shell-like debris~(orange) shown in Fig.~\ref{fig:orbits}. At points where the stellar mass loss is above the dotted grey line (i.e. $M_*/M_{*,\mathrm{z_{acc}}}=0.9$), satellites are considered to be intact. The dashed-dotted grey line denotes ``p300," the point where the satellite has lost all but $1/300$ of its total initial mass. The pericenter closest to this intersection is where the morphology metric is calculated in this analysis. If a satellite does not reach this point, its morphology is calculated at most recent pericenter.}
    \label{fig:mass_loss}
\end{figure}

There are several key differences between the simulations HJ15 used to define the morphology metric and \texttt{SatGen}. In particular, HJ15's morphology metric is defined by the dark matter properties of the satellite and focuses on satellite orbital properties; it does not directly account for the internal properties of the satellite. HJ15 does approximate the effect of stripping dark matter before stars by using $20\%$ of the initial satellite mass in the metric calculation, Eq.~\ref{mu}. Additionally, HJ15's morphology metric does not correct for dynamical friction and the variation in tidal radius as the satellite orbits. Dynamical friction and the tidal radius modeling are included in \SatGen\xspace and are therefore intrinsically included in the morphology metric for \texttt{StreamGen}. On the other hand, \SatGen\xspace does not track the location of the lost mass; it only reports the location of the satellite in the halo and its mass at each timestep. \SatGen\xspace's analytic mass loss and structural evolution prescriptions are described in Sec.~\ref{sec:SatGen}. HJ15 and \SatGen\xspace do not consider asphericity of halos. In this work, we adapt HJ15's morphology metric to classify the morphology of the tidal debris that would result from a given satellite generated by SatGen.
\\\\
\subsection{Sample StreamGen Runs}
\begin{figure}[t]
    \centering
    \includegraphics[width=0.9\columnwidth]{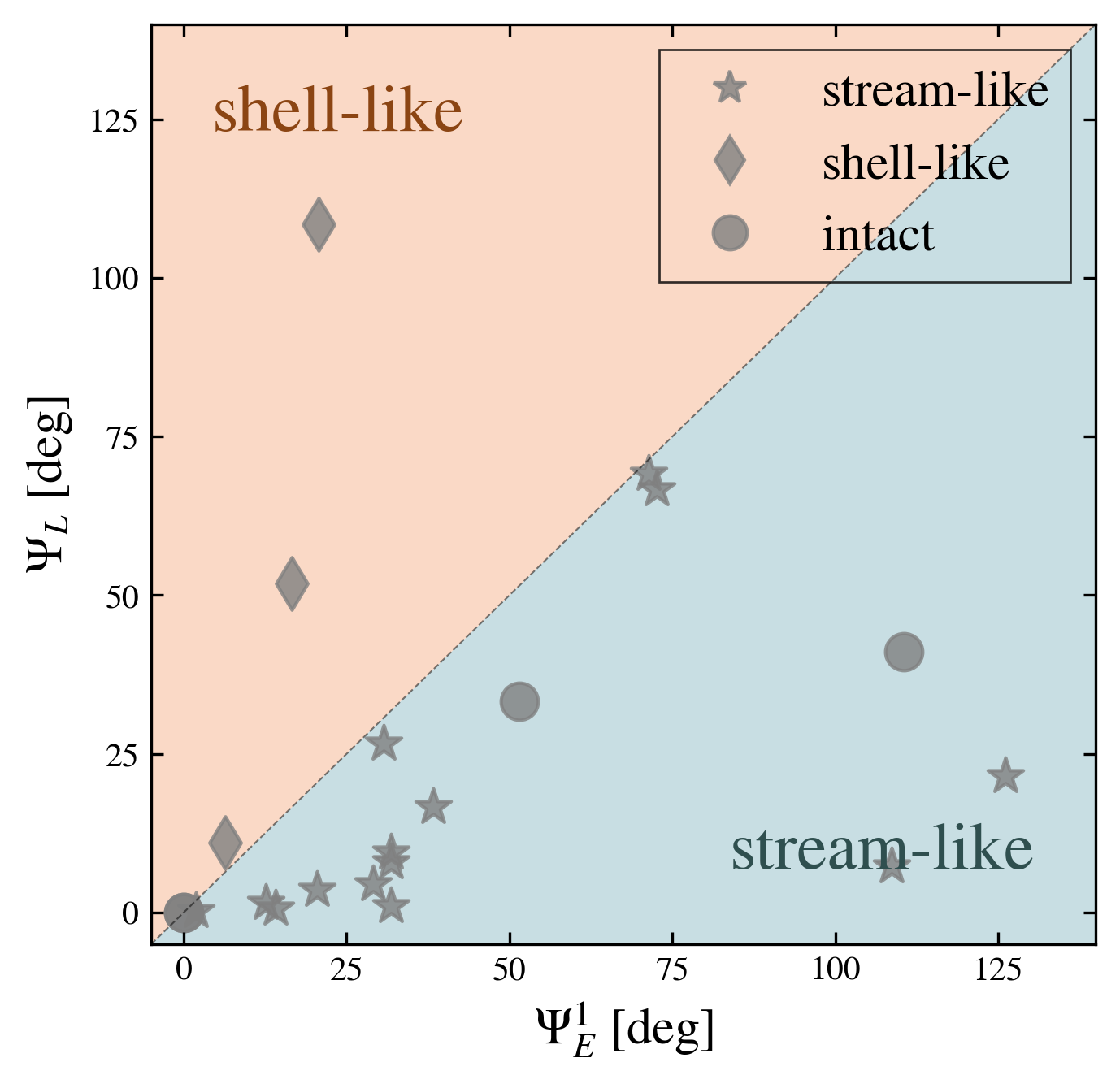}
    \caption{The 2D space spanned by the components of the morphology metric, the stream angle~($\Psi^1_E$) and the shell angle~($\Psi_L$), where $\Psi_L / \Psi^1_E$ is the morphology metric, Eq.~\ref{mu}. Satellite galaxies that fall below~(above) the grey, dashed 1:1 line have debris that is considered to be stream~(shell)-like and are depicted as stars~(diamonds). Satellites that have lost less than $10\%$ of their total stellar mass by $z=0$ are considered to be intact~(circles). These satellites belong to one example \SatGen\xspace galaxy with $M_{\mathrm{vir}} = 10^{11.98} M_\odot$.  }
    \label{fig:psiE_L_galaxy1}
\end{figure}
To give the reader a visualization of the \texttt{StreamGen} pipeline, we apply \texttt{StreamGen} to the two disrupting \SatGen\xspace satellite galaxies shown in Fig.~\ref{fig:orbits} and identified respectively by the morphology metric as stream-like~(top) and shell-like~(bottom). The stream has $M_\mathrm{vir} = 9.2 \times 10^9~M_\odot$ and $M_* = 4.2 \times 10^6~M_\odot$ at accretion and infalls with $(x,y,z) = (-51.9, -54.5, -86.8)$~kpc and $(v_x,v_y,v_z) = (193.0, -5.56, 91.4)$~km/s.  The shell has $M_\mathrm{vir} = 3.1 \times 10^9~M_\odot$ and $M_* = 1.6 \times 10^6~M_\odot$ at accretion and infalls with  $(x,y,z) = (12.7, 38.7, -101.3)$~kpc and $(v_x,v_y,v_z) = (-122.7, -182.0, 175.5)$~km/s. The dark, solid line in each panel of Fig.~\ref{fig:orbits} is the original \SatGen\xspace orbit from satellite accretion until p300, which is 8.6~(4.8)~Gyr for the stream~(shell), and the thin lines in the background are the orbits of the sampled particles. The eccentricity and period of the sampled orbits roughly match the original \SatGen\xspace orbit, with some spreading near apocenter. 

\begin{figure*}[ht]
    \centering
    \includegraphics[width=\textwidth]{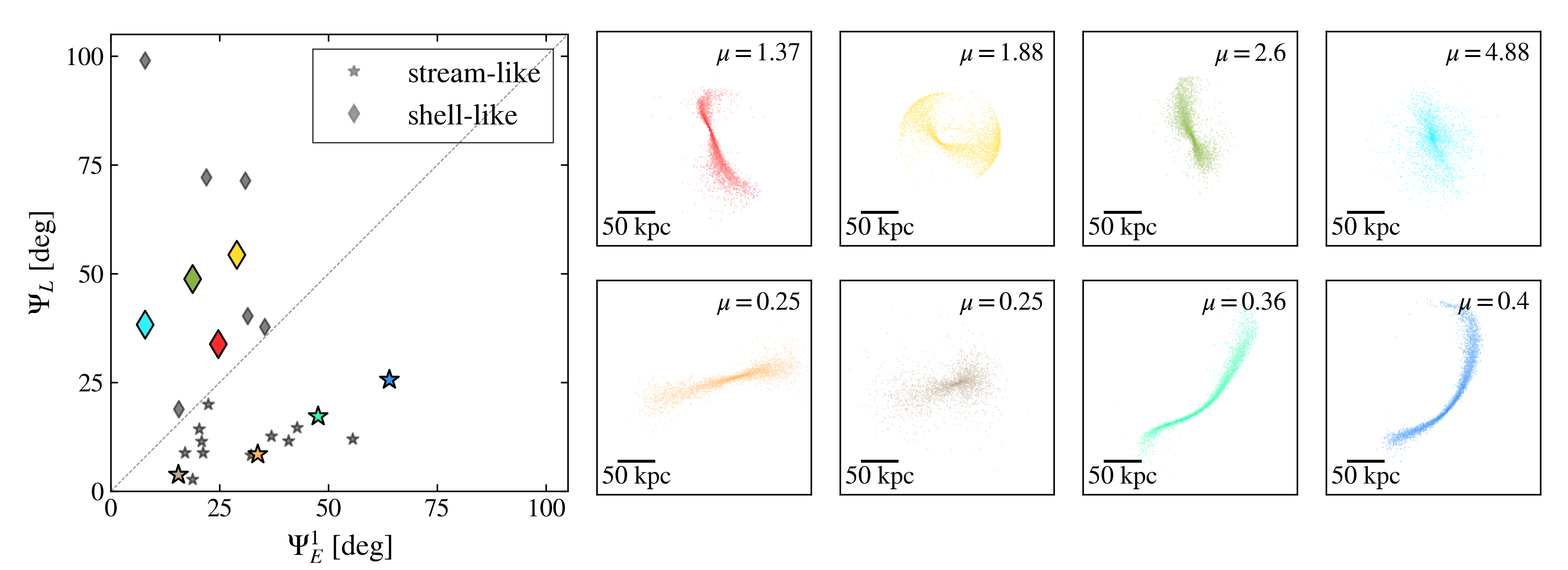}
    \caption{Application of the morphology metric to stellar debris generated by evolving an N-body satellite, with a total mass of $9.2 \times 10^9~M_\odot$ and stellar mass of $2.2 \times 10^6~M_\odot$, with varying orbits in an analytic host potential with a mass of $\roughly 5 \times 10^{11}~M_\odot$. The left-most plot shows the components of the morphology metric for each merger, where $\Psi_L / \Psi^1_E$ is the morphology metric, $\mu$. Disrupted satellite galaxies which fall below~(above) the grey, dashed 1:1 line are considered to be stream~(shell)-like. The colored points, a random sample of the disrupted satellite galaxies, have their corresponding debris plotted in the sub-figures on the right, viewed along the axis perpendicular to the orbital plane at most recent pericenter. The number in the upper-right corner of each sub-plot is the morphology metric. We verify that the morphology metric has a value of $\leq 1$ for debris which appears visually stream-like and $> 1$ for debris which appears visually shell-like. We purposefully include an example of debris~(brown) which is neither stream-like or shell-like but receives a classification as an example of the uncertainty from this effect in this analysis. See Appendix~\ref{app:uncertainty} for further discussion about how the classification uncertainty affects the final conclusions of this work.}
    \label{fig:validation}
\end{figure*}

Figure~\ref{fig:mass_loss} shows the corresponding fraction of dark matter~(solid) left to the fraction of stellar mass~(dashed) left in the progenitor of the stream~(blue) and shell~(orange) depicted in Fig.~\ref{fig:orbits}. Satellites begin to lose significant stellar mass when they have lost $\roughly 90\%$ of their dark matter mass, as expected from~\cite{Penarrubia_2008}. When a satellite has lost $<10\%$ of its initial stellar mass, it is considered intact~(i.e. have an intact, bright progenitor). This cut is somewhat arbitrary, but it is unlikely that tidal tails would be detectable if the satellite has only lost $10\%$ of its stellar mass~\citep{Penarrubia_2008,Shipp_2023}. Further, this cut is done as a post-processing step to \texttt{StreamGen}, so can be easily modified.  The point when 1/300 of the initial dark matter mass remains is marked by the grey dashed-dotted line. 

We also illustrate \texttt{StreamGen}'s ability to generate a population of stream-like and shell-like debris in Fig.~\ref{fig:psiE_L_galaxy1}. We plot $\Psi_L$ versus $\Psi^1_E$ for every satellite which has had a pericentric passage in an example \SatGen\xspace galaxy with $M_{\mathrm{vir}} = 10^{11.98}~M_\odot$. The dashed 1:1 line separates stream-like~(under the line) from shell-like~(over the line) morphologies. Stream-like debris~(shell-like debris) are indicated with star~(diamond) scatter points on this plot. Intact satellites are those which have lost less than $10\%$ of their peak stellar mass by redshift $z = 0$; they are indicated by circles.\footnote{Satellites which are on first infall are also considered to be intact in this analysis, but they do not have a pericenter and are therefore not shown in Fig.~\ref{fig:psiE_L_galaxy1}.} The greater abundance of stream-like debris than shell-like debris for this example host is similar to what we find for a larger sample of hosts, as we discuss in Sec.~\ref{sec:stream_props}.

\subsection{StreamGen Validation}\label{sec:validate_metric}

We next validate the procedure described in the previous section by running N-body simulations of satellite galaxy mergers, applying the \texttt{StreamGen} morphology metric. The goal is to test the implementation of the \texttt{StreamGen} pipeline by inspecting the debris from N-body simulations and assessing whether its visual morphology agrees with the metric's stream-like or shell-like classification. This validation tests the entire calculation of the \texttt{StreamGen} morphology metric assignment detailed in Sec.~\ref{sec:streamgen}, except we do not sample satellite particles, as we have full particle information in the N-body simulations, and we calculate the metric at most recent pericenter instead of p300. It is important to stress that we simulate similar galaxies but do not recreate exact \SatGen\xspace mergers. We test the \texttt{StreamGen} morphology metric's ability to accurately classify debris structures. 

Using \texttt{GALIC}~\citep{Yurin_2014}, we initialize a host galaxy with $M_\mathrm{vir} = 5.03 \times 10^{11}~M_\odot$ and $M_* = 2.29 \times 10^{10}~M_\odot$, as well as a massive, luminous satellite with $M_\mathrm{vir} =  9.2 \times 10^9~M_\odot$ and $M_* = 2.34 \times 10^6~M_\odot$. The satellite and host galaxies have dark matter halos and stellar bulges defined by Hernquist profiles; the host galaxy additionally contains an exponential stellar disk. \texttt{GALIC} creates a random N-body realization of the particle positions and iteratively adjusts the velocities until the galaxy stabilizes~(i.e. the code finds a stationary solution of the collisionless Boltzmann equation). Both the satellite and host have a stellar and dark matter particle resolution of $320~M_\odot$.  However, to minimize computation time, we use \texttt{agama} to create a multipole potential of the \texttt{GALIC} host~\citep{vasiliev2019agama} and merge the N-body satellite into it. To create a variety of different merger configurations, we scan over a number of initial positions and velocities for the satellite, where each coordinate has a value in our chosen range of $[-100, 100]$~kpc or km/s, respectively. We scan over the entire range to get a sample of stream-like and shell-like debris, but it is good to note that not all satellites within this range merge into the host. Then, we integrate the satellite orbit in the static host potential using \texttt{agama} for 13~Gyr. We apply the \texttt{StreamGen} morphology metric to the debris at its most recent pericenter, using $E_{\rm orb}$, $L_{\rm orb}$, $\Phi$, $r_\mathrm{apo}$, $r_\mathrm{peri}$, $T_r$, $\Delta \psi$, and $\alpha_r$, as defined in Sec.~\ref{morph_metric}.

We plot a selection of the results from these N-body simulations in Fig.~\ref{fig:validation}. The left-most plot is structured in the same way as Fig.~\ref{fig:psiE_L_galaxy1}. The eight plots to the right show a sample of different satellite mergers that yield different stellar debris configurations, color-coded to match points in the left-hand plot. Stellar debris is displayed at most recent pericenter, viewed along the axis perpendicular to the satellite's orbit at this point in time.    The number in the upper right-hand corner of each of the  smaller panels is the value of the morphology metric. Debris that visually appear stream-like have morphology metric values that are $< 1$. Stellar debris with the largest  metric values are clearly not stream-like.  There is an intermediate regime, where the morphology metric is $\roughly 1$, in which the debris could be considered to be neither stream- or shell-like, but somewhere in between. Finally, there exist some satellites, such as the satellite colored brown, which are identified to be stream-like or shell-like, but visually are difficult to classify. We investigate this uncertainty further in Appendix~\ref{app:uncertainty}.  We conclude that \texttt{StreamGen} can accurately separate very stream-like debris from debris that is distinctly not stream-like, with uncertainty in the intermediate regime. Appendix~\ref{app:uncertainty} explores the impact of the classification uncertainty on our final conclusions, demonstrating that the effect is negligible.  It is good to note that ``stream," ``shell," and even ``intact" are approximate boundaries; there is always inherent uncertainty in this classification, in simulations and observation. 
\\
\section{Results}
\label{sec:results}

This work aims to understand how the properties of the host galaxy affect the abundance and orbital distribution of tidal debris. We present populations of tidally-disrupting satellites across $\roughly 1500$ \SatGen\xspace host galaxies and discuss how the orbital properties of the satellites change as a function of host halo mass, disk mass, the ratio of disk scale radius to scale height, and baryonic feedback model. While it is likely that some of these modified parameters are correlated with each other, it is useful to gain intuition for how debris populations change as each of these host galaxy parameters is varied individually.
\begin{figure}[t]
    \centering
    \includegraphics[width=\columnwidth]{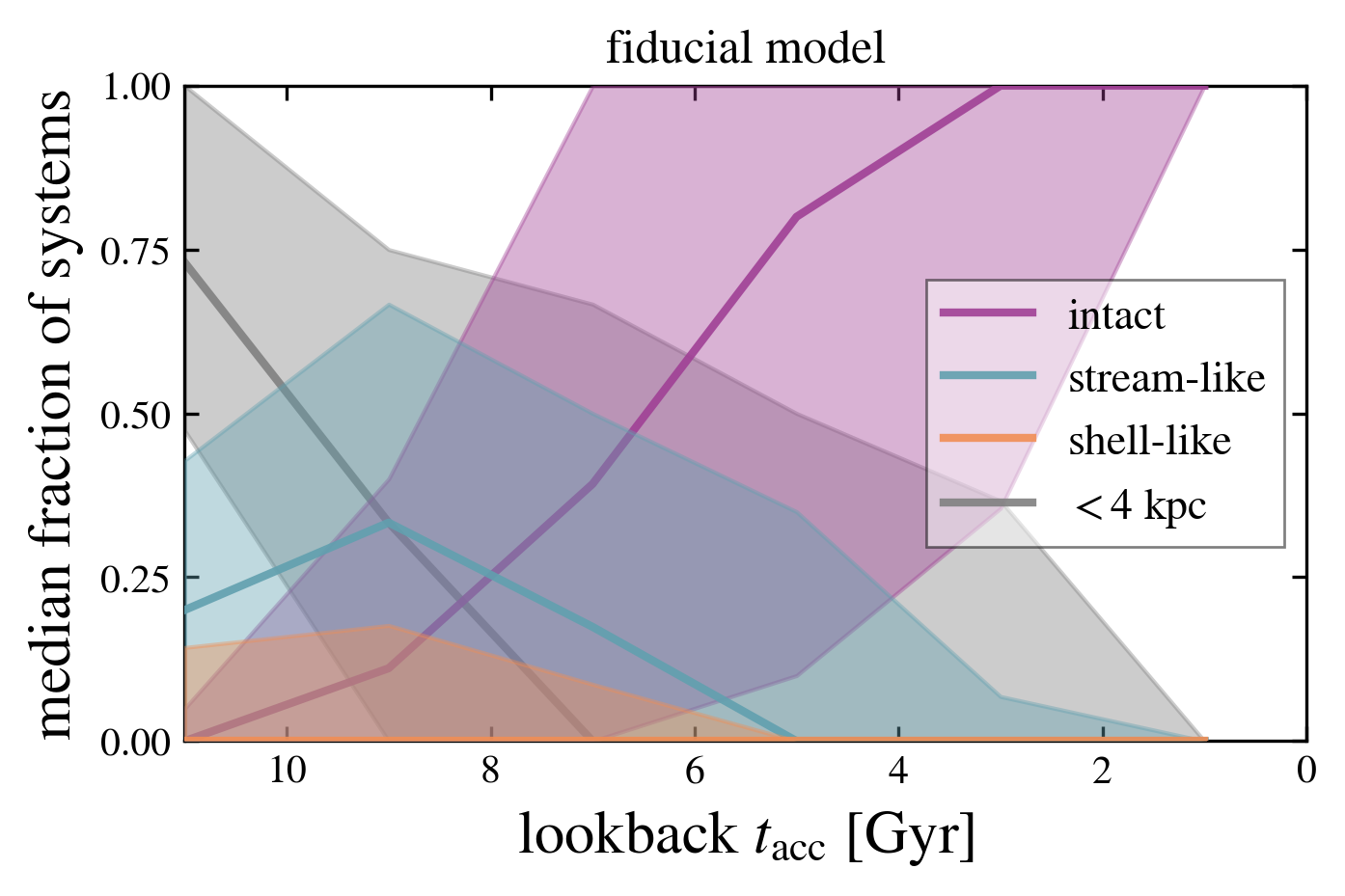}
    \caption{The fraction of satellites accreted at various lookback times for galaxies in the fiducial suite. The purple line denotes the median distribution for intact satellites. The blue~(orange) line represents the median distribution for stream-like~(shell-like) debris. The grey line indicates satellites with pericenters $< 4$ kpc from the center of the primary host and are not assigned a morphology metric value in this analysis. The corresponding shaded region indicates the $1\sigma$ scatter for each distribution. The debris behaves in predictable ways: satellites accreted at late times are largely intact today, while stream- and shell-like debris arise from satellites that accreted earlier. The grey band likely leads to shell-like debris.} 
    \label{fig:lookback}
\end{figure}
\begin{figure*}[ht]
    \centering
    \includegraphics[width=\textwidth]{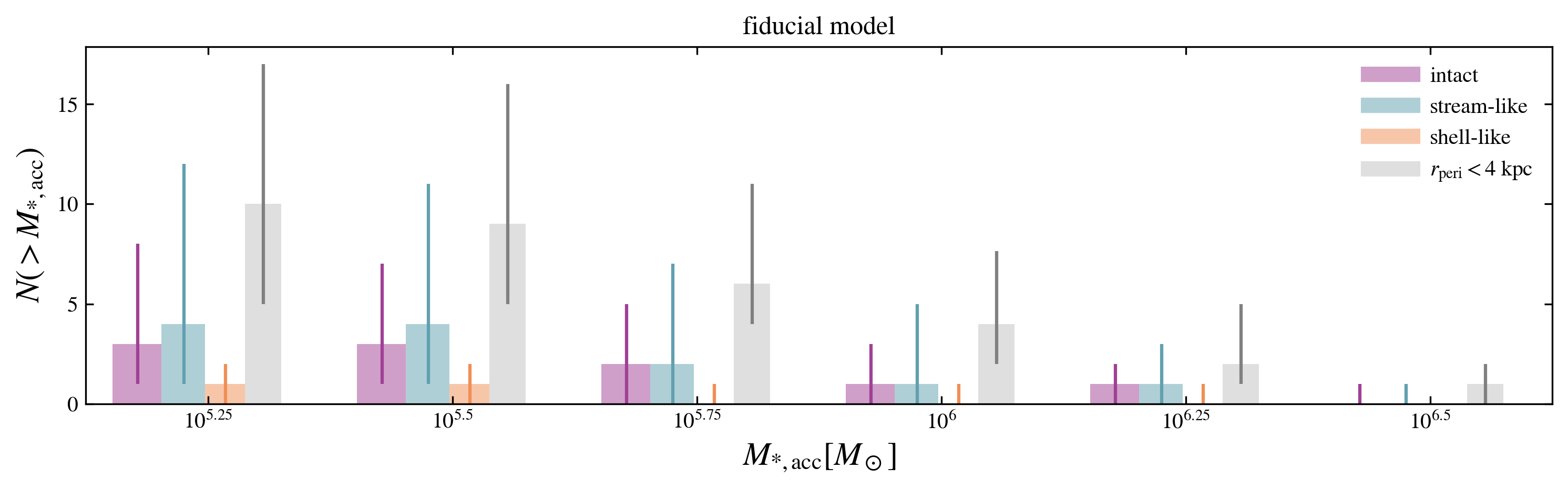} 
    \caption{The median cumulative stellar mass distributions of intact satellites~(purple), stream-like debris~(blue), and shell-like debris~(orange) across all host galaxies in the fiducial model. The grey bars show all of the satellites with pericenters less than 4~kpc. The vertical lines on each bar denote the $1\sigma$ scatter across all hosts. Halo-to-halo variance is clearly significant across all stellar masses and debris types.}
    \label{fig:mass_functions_1galaxy}
\end{figure*}

Current best estimates of the Milky Way's mass lie in the range of $[0.5$--$2.5] \times 10^{12}~M_\odot$~($[10^{11.7},10^{12.4}]$); see \cite{Wang_2020,Bobylev_2023} for recent reviews. The Milky Way's disk mass is estimated in the range of $[3.5 \pm 1] \times 10^{10}~M_\odot$, leading to a disk mass fraction that ranges from $[0.01,0.09]$~\citep{Bland_Hawthorn_2016}. According to \cite{Bland_Hawthorn_2016}, the scale height of the thin disk is $300\pm50$~pc, and the scale length is $2.6\pm0.5$~kpc, usually assuming that the density of the disk decreases as an exponential function of radius~(and height in the case of the smooth double exponential disk model).\footnote{The MN density profile for the disk used in this work is derived directly from the gravitational potential of the host, making it convenient for dynamical studies. However, this means that the scale radius and scale height calculated from observations are only approximate to the scale radius, $a$, and scale height, $b$, of the MN disk. Studying the effect of the specific disk model on tidal effects is worthwhile, but is not done in this work.} This places the range of the ratio of disk scale radius to disk scale height to be between $\sim6$--12.

As a baseline for comparison, we define a fiducial model with host galaxies that are approximately Milky Way-mass $M_{\mathrm{vir}} \in [10^{11.5}, 10^{12.5}]~M_\odot$, have a disk mass fraction of 0.05 times the host mass, a disk radius-to-height ratio of 25~(i.e.~quite flat), and a bursty model of feedback tuned to the NIHAO simulations. Note that in this work, we do not attempt to exactly model the Milky Way but rather investigate how variations in model parameters around Milky Way estimates can affect the population of tidal debris. Section~\ref{sec:stream_props} describes the substructure and orbital distribution for satellites in this fiducial model.  Then, Sec.~\ref{sec:vary_host_sim} shows how these abundances and distributions change as we vary the model parameters. 

\subsection{Fiducial Model} \label{sec:stream_props}

In this subsection, we examine the properties of intact satellites, as well as stream-like  and shell-like debris, across the fiducial suite.  To start, we  investigate the survival of satellite galaxies by considering the distribution of accretion times for intact satellites compared to those undergoing stream-like or shell-like disruption---see Fig.~\ref{fig:lookback}. The stream-like and shell-like debris included in this figure are identified at p300, the time in a satellite's orbit when the morphology metric is assigned. For the purposes of this analysis, we consider the morphology assigned at this point to be representative of the dominant structure of the debris at $z=0$, as described in Sec.~\ref{sec:streamgen}. For each separate population shown in Fig.~\ref{fig:lookback}, the solid line denotes the median across the different host galaxy distributions, evaluated at a given lookback time; the shaded region corresponds to the $1\sigma$ scatter of the distributions. The purple, blue, orange, and grey colors correspond respectively to intact satellites, stream-like debris,  shell-like debris, and  satellites that have a pericenter $< 4$~kpc. Satellites that are accreted at late times are typically still intact at $z=0$, as they have not had time to undergo significant disruption. Accordingly, those that result in stream-like debris at the present day typically accreted earlier, with the distributions peaking at a median value of $\sim 8$--9~Gyr. Shell-like structures are  subdominant at all lookback times. One of the key results of Fig.~\ref{fig:lookback}---which will be a major theme in the ensuing discussion---is the large spread observed for all populations, which highlights the significant halo-to-halo variance in debris morphology across the \SatGen\xspace galaxies. 
\begin{figure*}[ht]
    \centering
    {\includegraphics[width=\textwidth]{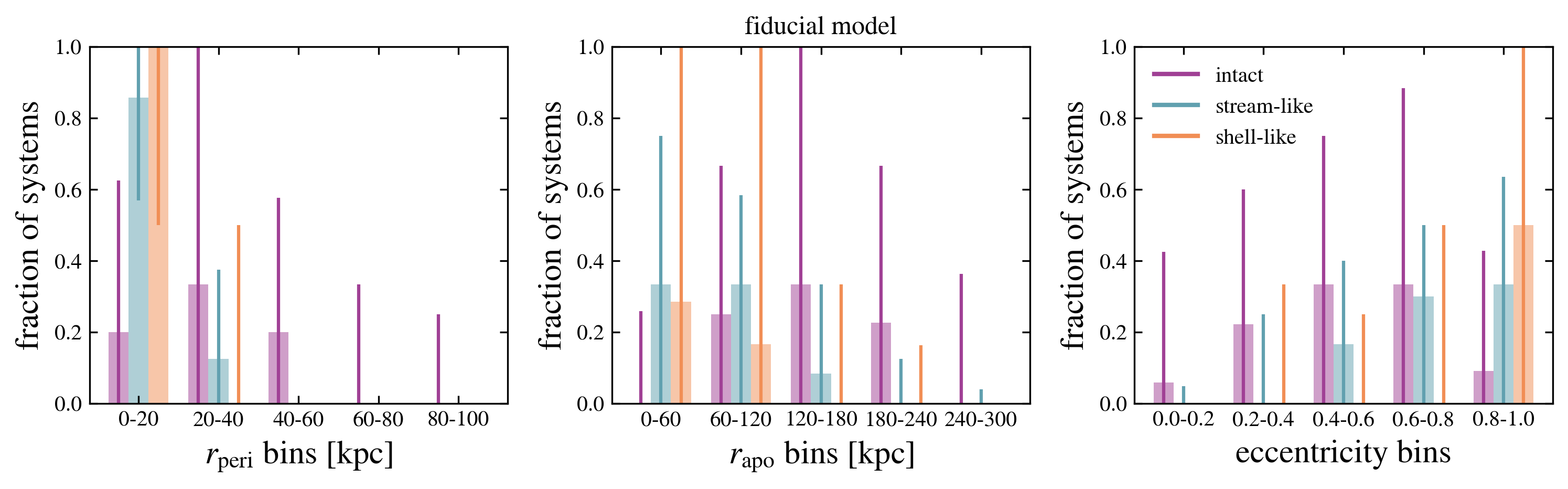}}
    \caption{The median distributions of $r_\mathrm{peri}$, $r_\mathrm{apo}$, and eccentricity for intact satellites~(purple), stream-like debris~(blue), and shell-like debris~(orange) for all galaxies in the fiducial model. The vertical line on each bar shows the $1\sigma$ spread across all galaxies. Stream-like and shell-like debris are largely associated with satellites on high-eccentricity orbits with pericenters of 0--20~kpc and apocenters in the range of 0--120~kpc.}
    \label{fig:peri_apo_hist_1galaxy}
\end{figure*}

We emphasize that the shell-like and stream-like debris included in Fig.~\ref{fig:lookback} have progenitor satellites with pericenters greater than 4~kpc. Recall that this cut is put in place to mitigate numerical issues for orbits passing near the center of the host. Satellites with pericenters smaller than 4~kpc  will be significantly disrupted by strong tidal forces in the inner regions of the halo and are unlikely to form coherent streams~\citep{garrisonkimmel2017}. We do not explicitly calculate the morphologies of these satellites, so we categorize them separately, but because of intense tidal processing in the inner regions of the halo, we expect most to be shell-like or fully phase mixed~\citep{Johnston_2016}. The results from  Fig.~\ref{fig:lookback} (grey line/band) suggest that there may be additional contributions to the \emph{total number} of shell-like debris (and possibly some stream-like debris) in each galaxy of the suite, beyond what is included in our shell-like and stream-like classification.

The stellar mass distribution function of satellite galaxies  provides insights into satellite survival rates and a point of comparison between observations and simulations. The bar charts in Fig.~\ref{fig:mass_functions_1galaxy} show the median cumulative peak stellar mass distribution of satellites which remain intact~(purple), as well as those which become stream-like~(blue) and shell-like~(orange). The vertical lines on each bar indicate the $1\sigma$ scatter around the median for the galaxies in the fiducial model---emphasizing the significant halo-to-halo variance. For masses below $M_{*,\mathrm{acc}} \sim 10^{5.75}~M_\odot$, there is  only a slight increase in the number of satellites that yield stream-like debris to those that remain intact. The number of satellites yielding shell-like debris is subdominant at all stellar masses. 
The grey bars show the cumulative stellar mass distribution for satellites with a pericenter $< 4$~kpc. The contribution from these satellites dominates across all stellar masses. As shown in Fig.~\ref{fig:lookback}, this debris is primarily from the large fraction of early-infalling satellites, which are expected to contribute to the phase-mixed component of the host~\citep{Johnston_2016}. 
\begin{figure*}[ht]
    \centering
    \includegraphics[width=0.32\textwidth]{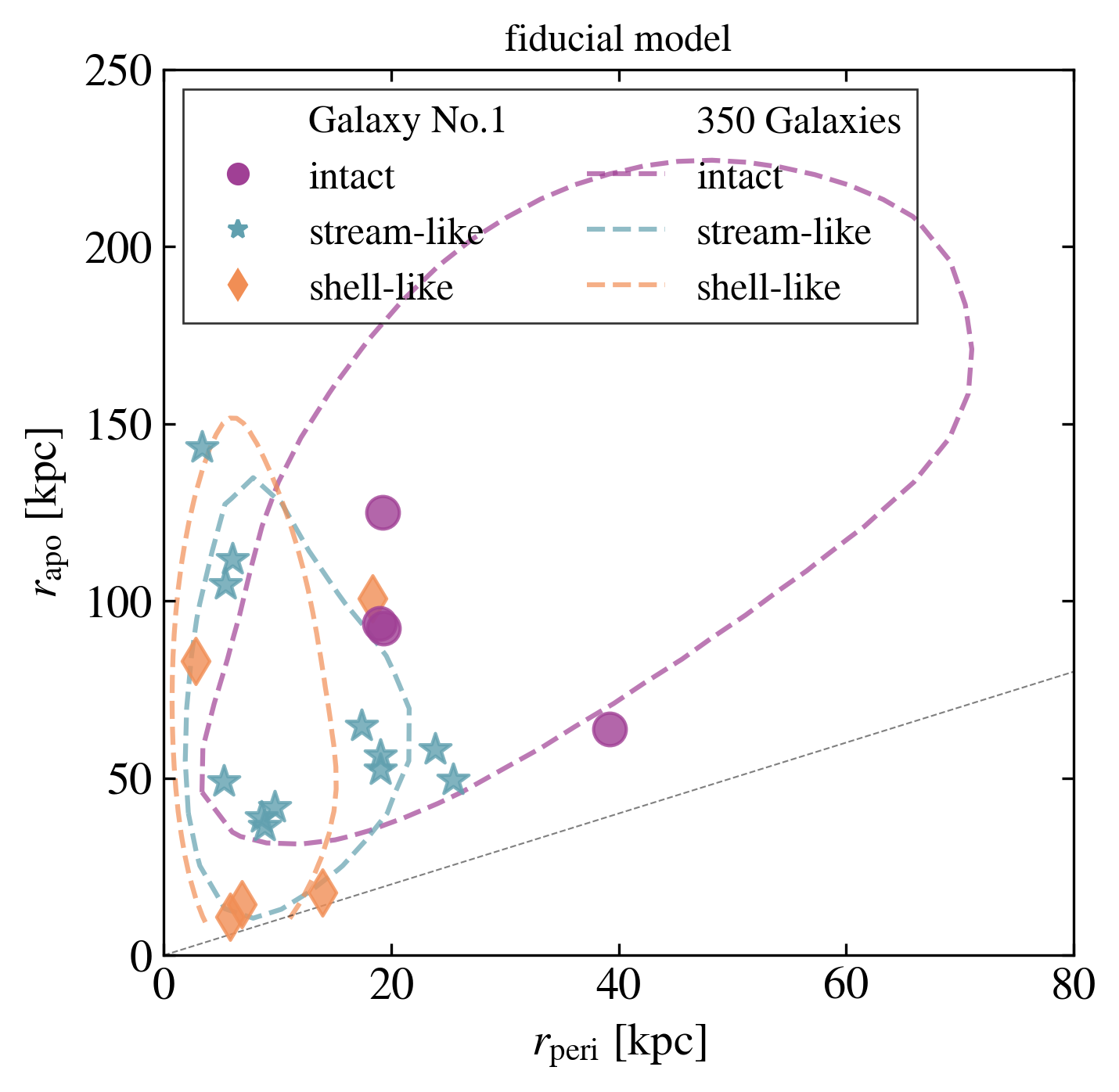}
    \includegraphics[width=0.32\textwidth]{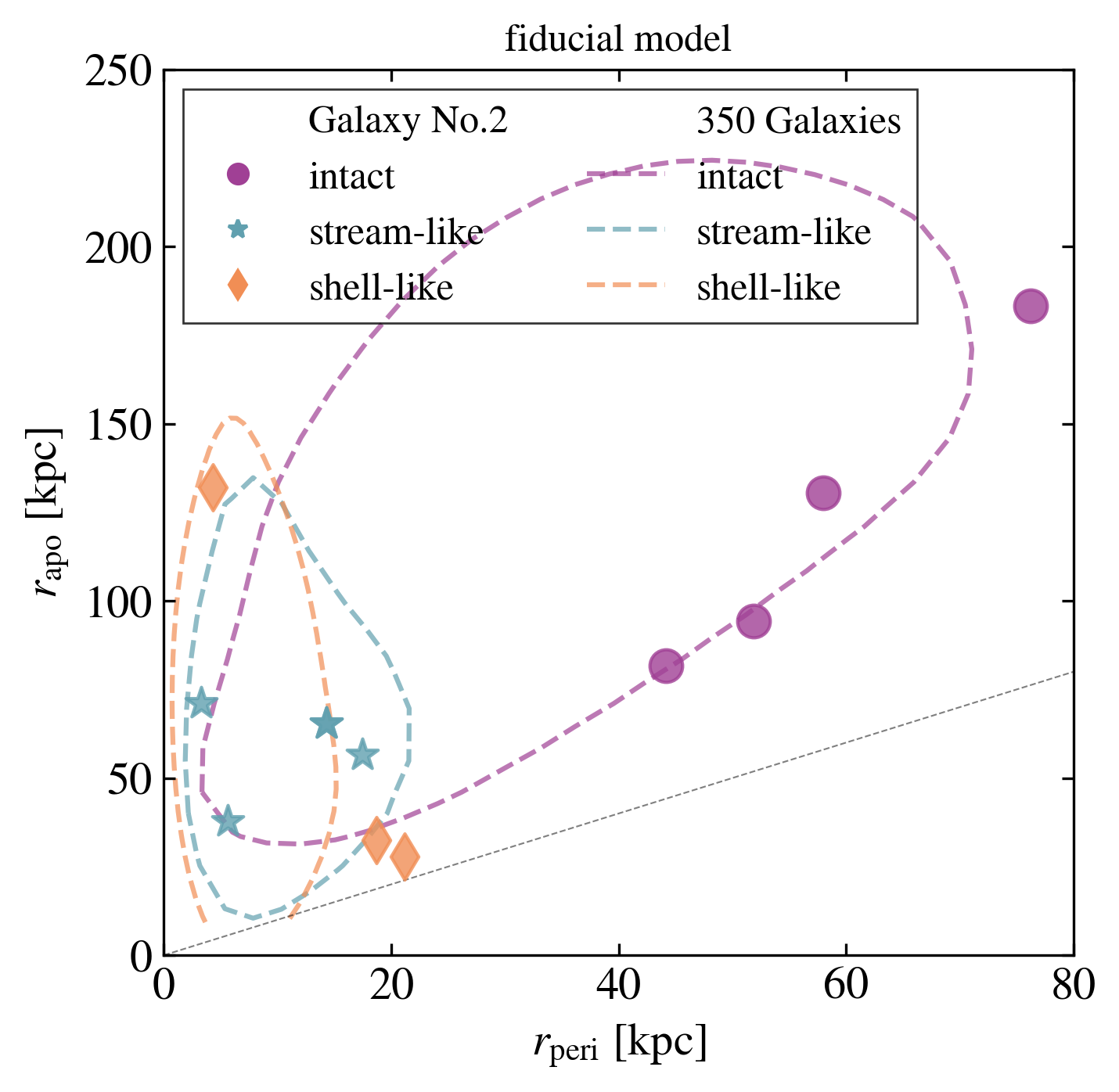}
    \includegraphics[width=0.32\textwidth]{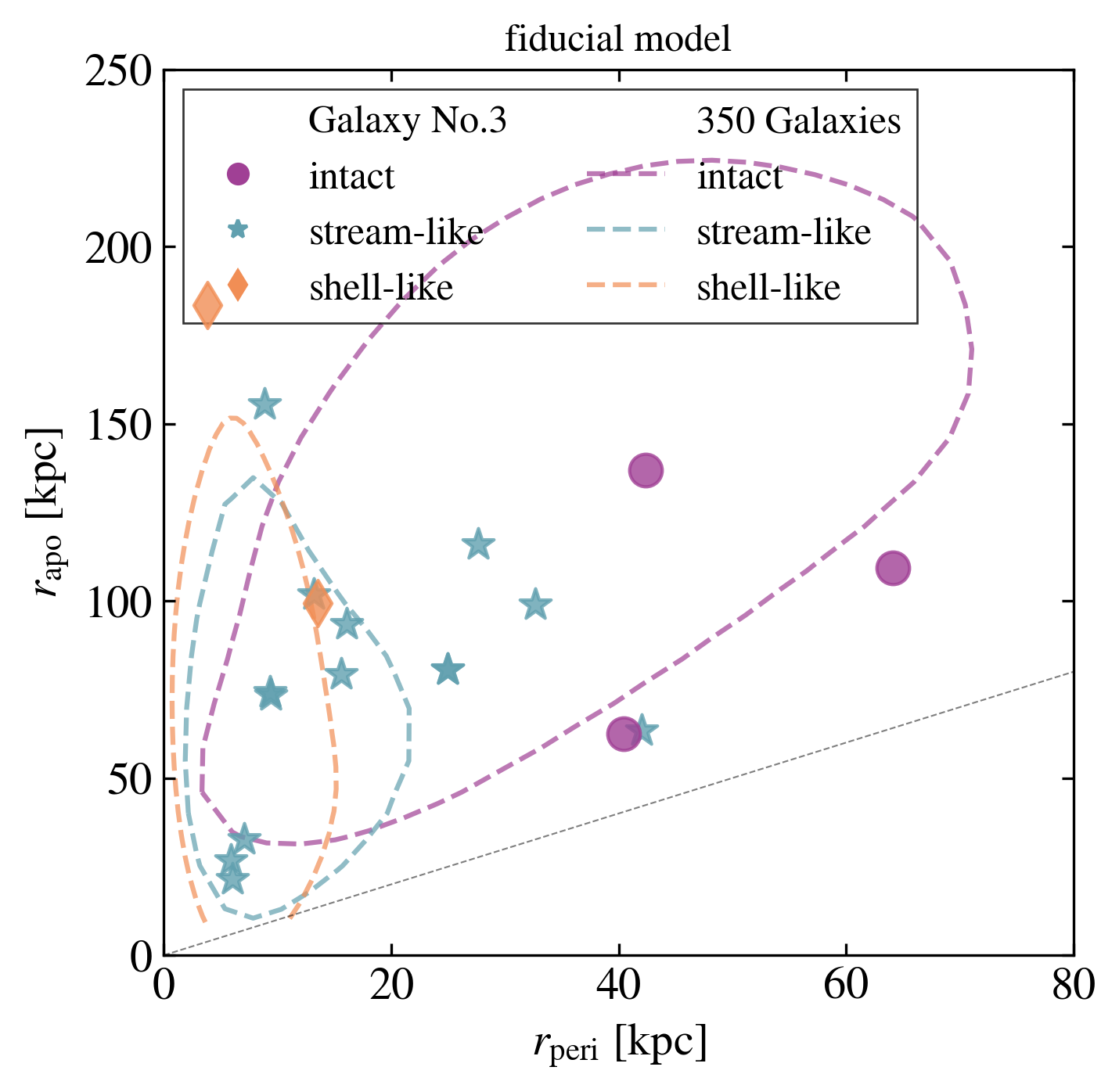}
    \caption{Halo-to-halo scatter in the distribution of tidal debris. Each panel shows the distribution of tidal debris in pericenter-apocenter space for a different host galaxy with $M_\mathrm{vir} = 10^{11.98}~M_\odot$. The scatter points correspond to stream-like debris~(blue stars), shell-like debris~(orange diamonds), and intact satellites~(purple circles). The dashed contours are the same in each panel and indicate the $1\sigma$ scatter for all satellites across all galaxies in the fiducial model.  The different colors correspond to intact satellites~(dashed purple), stream-like debris~(dashed blue), and shell-like debris~(dashed orange). There is significant variation in the distribution of tidal debris between galaxies of the same $M_\mathrm{vir}$.}
    \label{fig:peri_apo_1galaxy}
\end{figure*}

Figure~\ref{fig:peri_apo_hist_1galaxy} shows the distributions of $r_\mathrm{peri}$, $r_\mathrm{apo}$, and eccentricity for intact satellites~(purple), stream-like debris~(blue), and shell-like debris~(orange), calculated from the reintegrated satellite orbitsc\footnote{The values for $r_\mathrm{peri}$ and $r_\mathrm{apo}$ come from the median reintegrated orbit of the satellite at p300, as explained in Sec.~\ref{sec:streamgen}. These values should converge to the original \texttt{SatGen} $r_\mathrm{peri}$ and $r_\mathrm{apo}$ values, but it is possible that the pericenter of the debris could extend below $4$~kpc due to this reintegration. 
}h bar corresponds to the median fraction of each system that falls into the respective bin on the horizontal axis. The vertical line on each bar represents the $1\sigma$ scatter of the respective distributions across all hosts in the fiducial model. Satellites that yield shell-like debris have the lowest-peaking $r_\mathrm{peri}$ and $r_\mathrm{apo}$, followed by those that yield  stream-like debris, and then the intact satellites themselves. The stream-like and shell-like debris have pericenters that are almost entirely contained within 0--20~kpc and  apocenters concentrated in the range from 0--120~kpc. Satellites that yield either stream-like or shell-like debris have eccentricities peaked at values close to 1, although the former has a longer tail towards smaller values.  The eccentricity distribution for intact satellites is peaked around $\sim0.6$. These results are consistent with literature that suggests that satellites on eccentric orbits typically lead to tidal tails~\citep{Read_2006,garrisonkimmel2017,sawala_2017,Shipp_2018, Piatti_2020,Green_2021_disc,yoon2024}. There is significant variance over tidal debris $r_\mathrm{peri}$, $r_\mathrm{apo}$, and eccentricity distributions between galaxies.  

In Fig.~\ref{fig:peri_apo_1galaxy}, we examine the 2D distribution of the median re-integrated $r_\mathrm{peri}$ and $r_\mathrm{apo}$ for the tidal debris, which will provide a point of comparison with existing literature~\citep{Shipp_2023} To demonstrate the effect of the halo-to-halo variance on the debris between host galaxies of the same mass, each panel in Fig.~\ref{fig:peri_apo_1galaxy} shows the scatter for a different host galaxy with $M_\mathrm{vir} = 10^{11.98}~M_\odot$. The purple circles represent intact satellites, the blue stars represent stream-like debris, and the orange diamonds represent shell-like debris.  The dashed contours show $1\sigma$ scatter, calculated using the \texttt{scipy} gaussian kernel density estimate, for all the satellites across all the galaxies in the fiducial model. As expected, the intact satellites exist at higher $r_\mathrm{peri}$ and $r_\mathrm{apo}$, while stream-like debris and shell-like debris are scattered at lower $r_\mathrm{peri}$ and $r_\mathrm{apo}$. Shell-like debris tend to extend to higher $r_\mathrm{apo}$ and lower $r_\mathrm{peri}$ because of their highly eccentric orbits.
\begin{table}
    \begin{center}
    \centering
    \footnotesize
    \renewcommand{\arraystretch}{2}
    \noindent
    \begin{tabular}{ C{1.5 cm}|C{1.5 cm} C{1.5 cm} C{1 cm} C{1 cm}}
    \Xhline{3\arrayrulewidth}
     Model & No. Galaxies & Feedback \newline Model & $f_d$ & $a$/$b$ \\
      \hline
      fiducial model & 370 & NIHAO & 0.05 & 25\\
      puffier disk & 367 & NIHAO & 0.05 & 11\\
      more-massive disk & 347 & NIHAO & 0.1 & 25\\
      smoother feedback & 434 & APOSTLE & 0.05 & 25\\
     \Xhline{3\arrayrulewidth}
    \end{tabular}
    \caption{We generate merger trees each with NIHAO and APOSTLE feedback models. Our fiducial \SatGen\xspace galaxy suite includes galaxies that  resemble the Milky-Way mass galaxies in the FIRE simulation; we evolve the NIHAO feedback merger trees and impose that the host have a 0.05 disk mass fraction~($f_d$) (i.e. the fraction of the total dark matter mass that is the disk) and a disk scale radius-to-scale height~($a$/$b$) ratio of 25. We evolve the APOSTLE galaxies under the same conditions. We create galaxies from the NIHAO merger trees which match the fiducial model, but imposing a 0.1 host disk fraction, as well as additional galaxies which also match the fiducial model but a disk scale radius-to-scale height ratio of 11, to match recent estimates~\citep{Yan_2019}. The total number of galaxies per model reflects the number which were processed in the allotted computation time period.} 
    \label{table:models}
    \end{center}
\end{table}

\begin{figure*}[ht]
    \centering
    \includegraphics[width=0.34\textwidth]{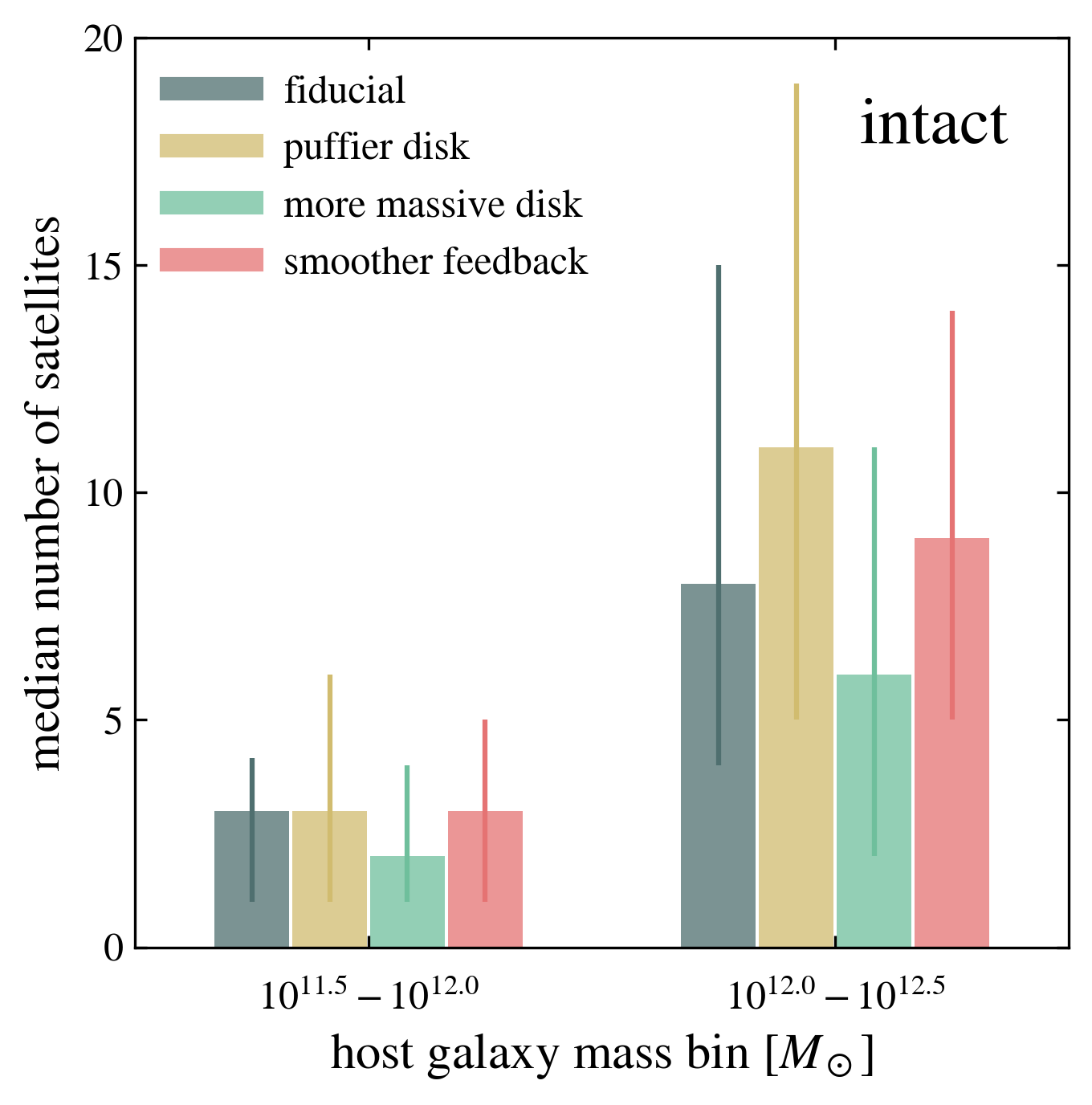}
    \includegraphics[width=0.34\textwidth]{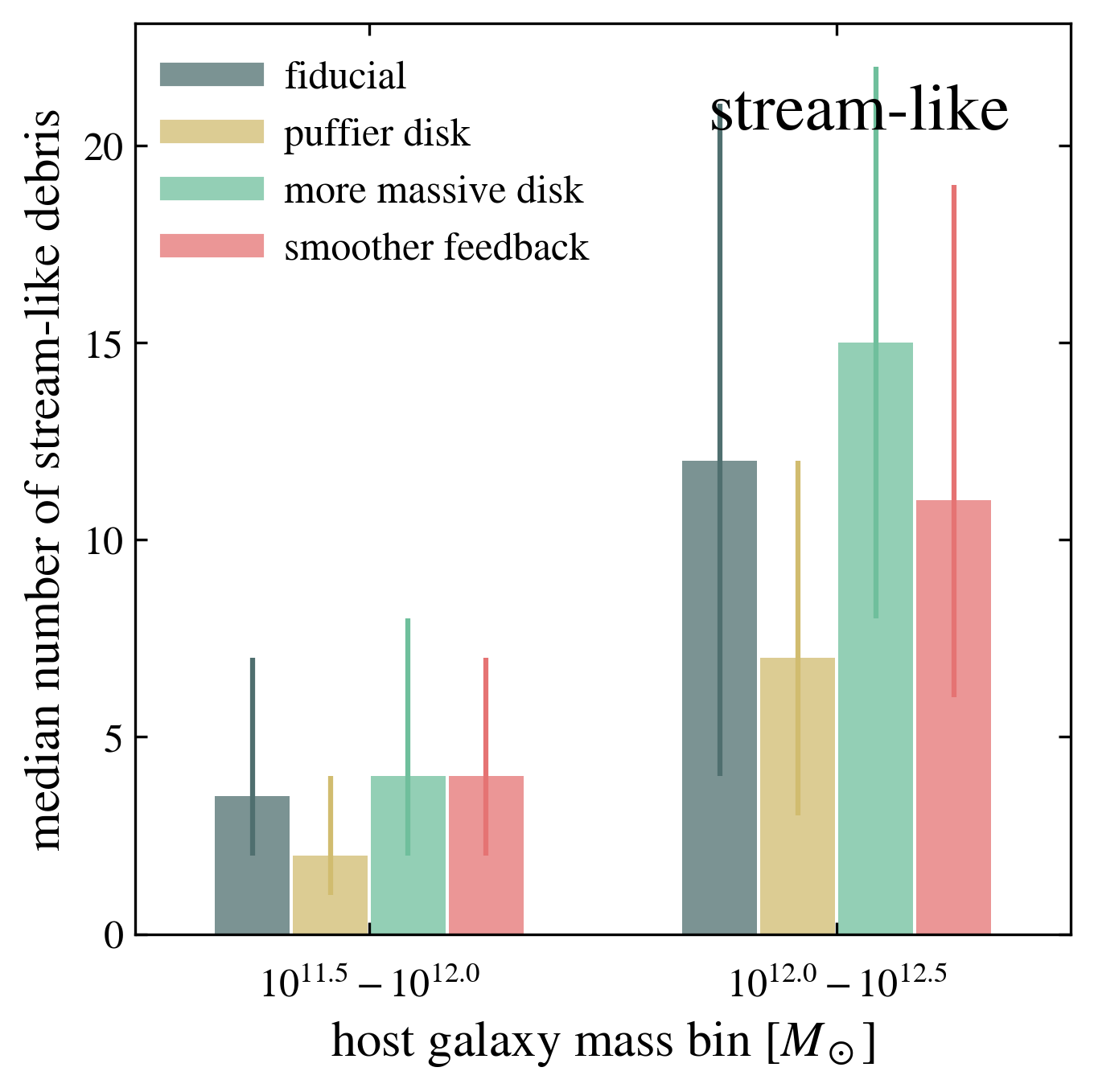}
    \includegraphics[width=0.34\textwidth]{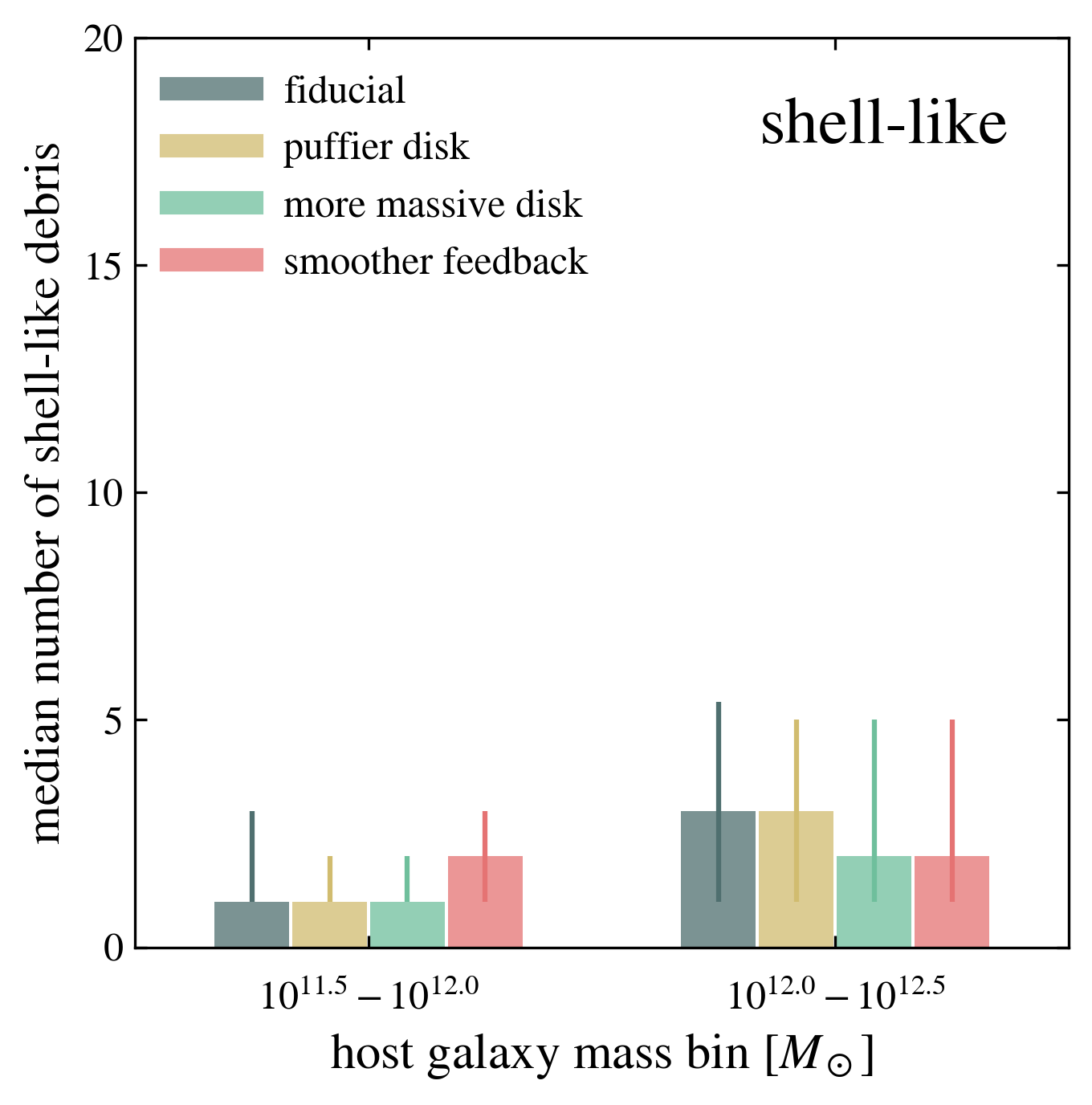} 
    \includegraphics[width=0.34\textwidth]{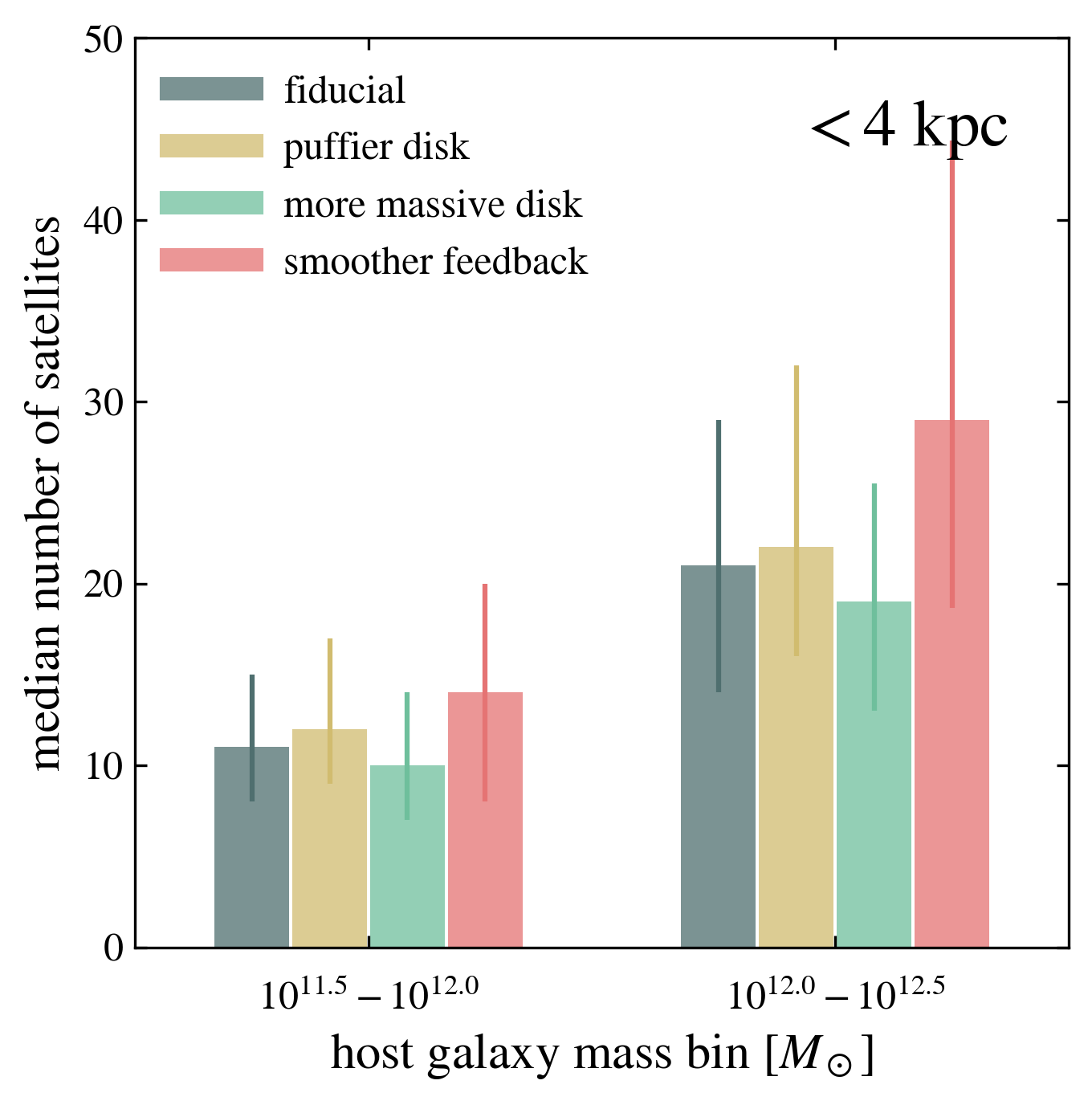} 
\caption{The median number of satellites for low-mass and high-mass host galaxies, across all models: the fiducial model~(black), the more-massive-disk model~(green), the puffier-disk model~(yellow), and the smoother-feedback model~(red). The vertical line on each bar indicates the $1\sigma$ spread on the median across all the galaxies in a given suite. The upper-left panel shows the distribution of intact satellites; the upper-right panel shows stream-like debris; the bottom-left panel shows shell-like debris; the bottom-right panel shows debris with $r_{\rm peri} < 4$~kpc. The numbers of each type of debris are consistent within halo-to-halo variance across models for a given range of host masses. However, there are still relevant trends between the median abundances of each type of tidal debris. For example, there are more intact satellites and fewer streams in the puffier-disk model. There are fewer intact satellites and more streams in the more-massive-disk model. As discussed further in Sec.~\ref{sec:apostle}, the stream- and shell-like debris are relatively insensitive to the choice of feedback model due to the implementation of the morphology metric in \texttt{StreamGen}. }
    \label{fig:numbers_grid}
\end{figure*}
\cite{Shipp_2023} performed a detailed study of pericenter and apocenter distributions for stream-like debris in the FIRE simulations~\citep{Hopkins_2015,Hopkins_2018, Hopkins_2022,10.1093/mnras/staa2101}.  They found a potential discrepancy between simulations and observations: the stream-like debris identified in 13 Milky Way analogs consistently had higher $r_\mathrm{peri}$ and $r_\mathrm{apo}$ than observed streams in the Milky Way. \texttt{StreamGen} provides an opportunity to compare these results against a larger sample of semi-analytically modelled Milky Way hosts.  Our fiducial model is roughly FIRE-like in terms of host mass range: the fiducial model hosts have $M_\mathrm{vir} \in [10^{11.5},10^{12.5}]$, and the FIRE galaxies exist in the mid-high end of this range. FIRE galaxies have thin disks with scale lengths in the range of 3--5~kpc and scale heights around 200--500~pc~\citep{Ma2017, ElBadry2018,Sanderson_2020}. This places the range of the ratio of disk scale radius to scale height to be between $\roughly 6$--$15$. The fiducial model has a flatter disk than the FIRE galaxies, with a disk scale radius to scale height of 25. Finally, the fiducial model has feedback tuned to the burstier feedback of the NIHAO simulations~\citep{10.1093/mnras/stv2856,10.1093/mnras/staa2790}, characterized by intermittent, intense supernovae outflows. This burstier feedback model is similar---though not the same in detail---to the FIRE feedback prescription. Selecting for satellites with $M_* > 5 \times 10^5~M_\odot$ as done in \cite{Shipp_2023}, we find good agreement with the abundance of \texttt{StreamGen} stream-like debris in the fiducial model to the FIRE stream-like debris, especially considering the halo-to-halo variance. However, we find stream-like debris that exist at $r_\mathrm{apo}< 50$~kpc and $r_\mathrm{peri}< 20$~kpc, lower than what was found in FIRE.  Whether or not these differences could be attributed to differences the modeling of host properties will be discussed in Sec.~\ref{sec:conclusion}.\\

\subsection{Variations of Host Properties and Feedback}\label{sec:vary_host_sim} 

\begin{figure*}[ht]
    \centering
    \includegraphics[width=0.315\textwidth]{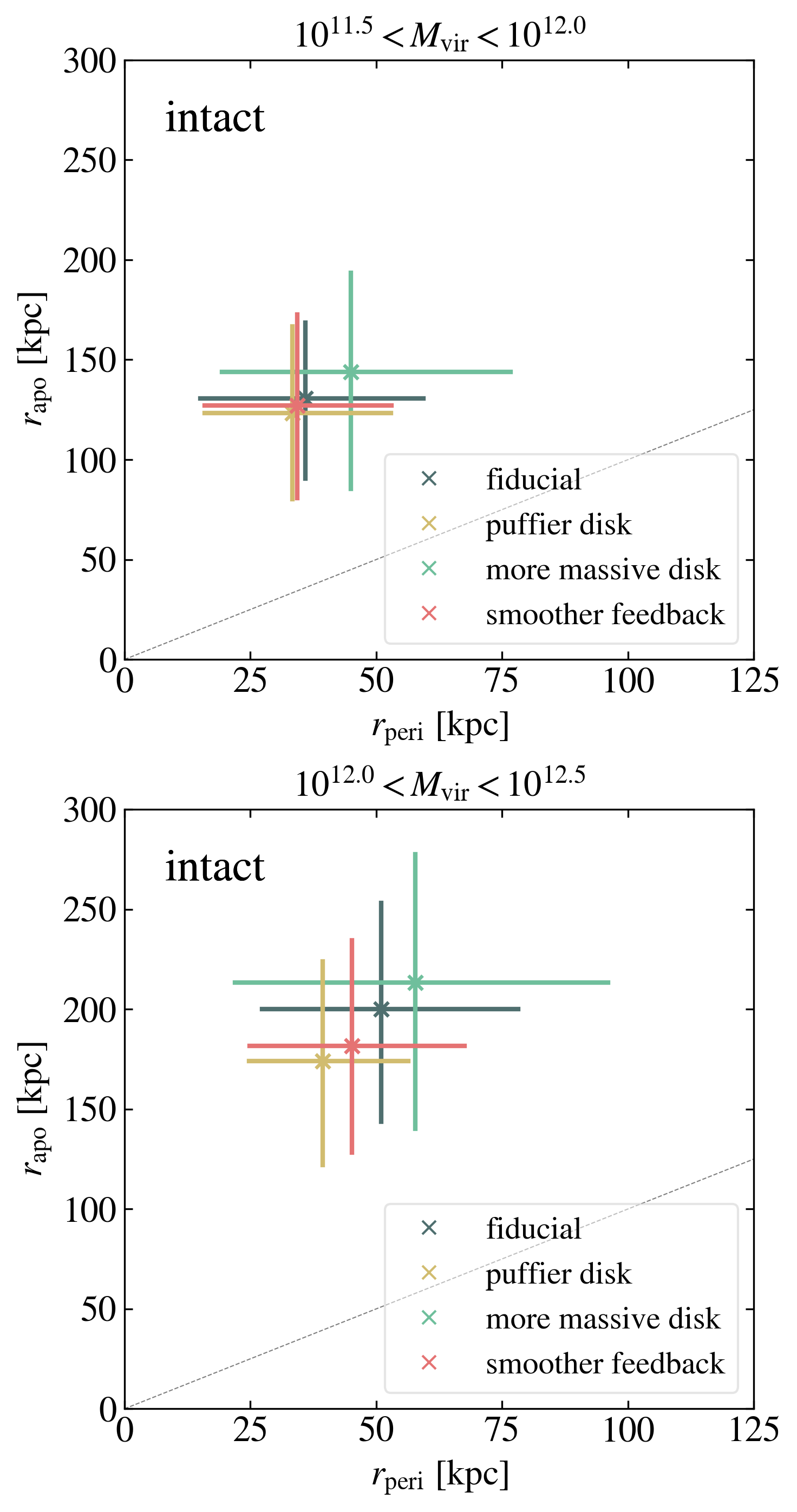}
    \includegraphics[width=0.31\textwidth]{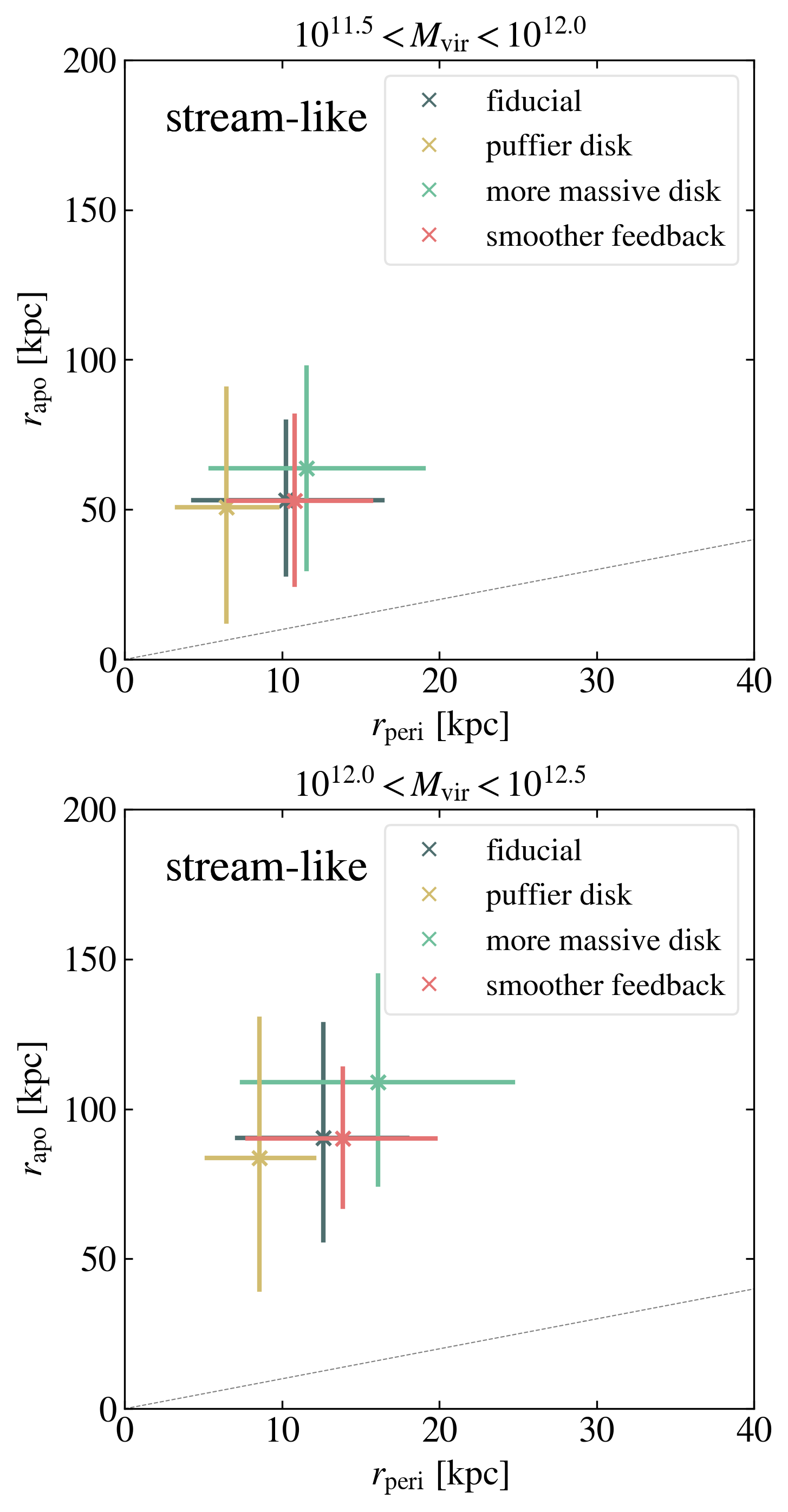}
    \includegraphics[width=0.31\textwidth]{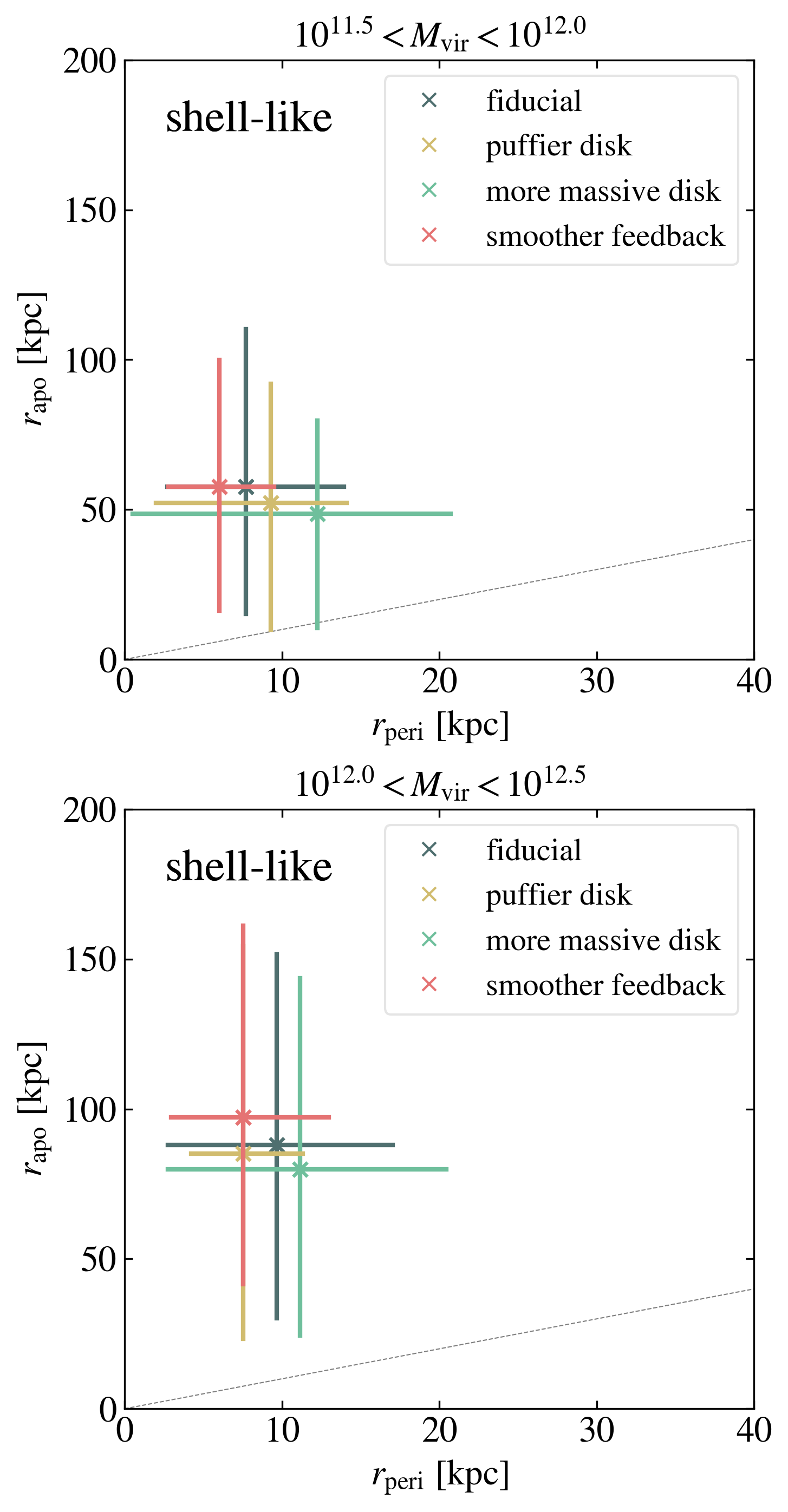} 
    \caption{The pericenter-apocenter distributions for intact satellites~(first column), stream-like debris~(second column), and shell-like debris~(third column). The top row shows results for satellites and tidal debris in low-mass~($M_{\mathrm{vir}} \in [10^{11.5}, 10^{12.0}]~M_\odot$) host galaxies, and the bottom row shows the results for satellites and tidal debris in high-mass~($M_{\mathrm{vir}} \in [10^{12.0}, 10^{12.5}]~M_\odot$) galaxies. The black lines indicate the fiducial model; the green lines indicate the more-massive-disk model; the yellow lines indicate the puffier-disk model, and the red lines indicate the smoother-feedback model. Each `$\times$' is centered on the median of these values across all the galaxies in the suite, with the horizontal and vertical lines denoting the $1\sigma$ scatter (in both $r_\mathrm{peri}$ and $r_\mathrm{apo}$) across all galaxies. The thin, grey dashed line is the 1:1 line indicating circular orbits. Halo-to-halo variance dominates any differences in the distributions due to model differences. However, the relative shifts in these distributions between models can be explained using the analytic prescriptions for tidal stripping and morphology classification. The shell-like distributions should be interpreted with care given the small-number statistics of this sample. As discussed further in Sec.~\ref{sec:apostle}, the stream- and shell-like debris are relatively insensitive to the choice of feedback model due to the implementation of the morphology metric in \texttt{StreamGen}. Note the different axes for the intact satellites.}
    \label{fig:peri_apo_grid}
\end{figure*}

This subsection considers variations on the fiducial model with a puffier disk, a more massive disk, and a smoother model of baryonic feedback. These models are described sequentially in this subsection and are summarized in Table~\ref{table:models}. We find that halo-to-halo variance is the dominant effect on the distributions, but that there are still relevant trends between the separate cases. Throughout, we use two sets of figures to compare across models.  To orient the reader, we first present these for the fiducial model and then generalize from there.  
The bar charts in Fig.~\ref{fig:numbers_grid} present the median number of satellites for each model for two ranges of host galaxy mass: low mass~$M_\mathrm{vir} \in [10^{11.5},10^{12.0}]~M_\odot$ and high mass~$M_\mathrm{vir} \in [10^{12.0},10^{12.5}]~M_\odot$. The upper-left panel shows the median number of intact satellites; the upper-right panel shows the stream-like debris; the bottom-left panel shows the shell-like debris; and the bottom-right panel shows satellites with $r_{\rm peri} <4$~kpc.  The vertical line on each bar corresponds to the $1\sigma$ spread on the median across all galaxies in the suite. The results for the fiducial model are indicated in black. There are, on average, over twice as many intact and disrupting satellites in the high-mass host galaxies than low-mass hosts.  As a consistency check for the intact satellites, our predictions for the number of intact satellites are consistent with the ELVES survey of Local Volume hosts (Fig.~1 of \cite{Danieli_2023}) when comparing in the same host galaxy mass range. The other bars correspond to models that will be discussed below.

Figure~\ref{fig:peri_apo_grid} shows the pericenter-apocenter distributions for the intact satellites~(left column), stream-like debris~(middle column), and shell-like debris~(right column) across models. 
The top~(bottom) row  shows results for low~(high)-mass host galaxies. The dashed, black 1:1 line in each panel indicates circular orbits. For each galaxy in a given suite, we calculate the median $r_\mathrm{peri}$ and $r_\mathrm{apo}$ for each specific debris type. In Fig.~\ref{fig:peri_apo_grid}, each `$\times$' is centered on the median of these values across all the galaxies in the suite, with the horizontal and vertical lines denoting the $1\sigma$ scatter (in both $r_\mathrm{peri}$ and $r_\mathrm{apo}$) across all galaxies. The halo-to-halo variance is significant for each model. While we present the results for the shell-like debris, we caution that these are statistics limited (as shown in Fig.~\ref{fig:numbers_grid}) and thus do not summarize them in the ensuing discussion.

Within each mass range of Fig.~\ref{fig:peri_apo_grid}, the intact satellites exist at higher $r_\mathrm{peri}$ and $r_\mathrm{apo}$ than the stream-like debris for the fiducial model (shown in black). The stream-like debris is shifted towards more eccentric orbits, as we would expect given the results from Fig.~\ref{fig:peri_apo_1galaxy}. Lastly, we note that there is an overall scaling with larger host galaxy mass to larger pericenters and apocenters. For the fiducial model, the intact satellite centroid shifts by $\Delta(r_\mathrm{peri}, r_\mathrm{apo}) = (+15, +70)$~kpc between the high-mass and low-mass group, and the stream-like debris shifts by $\Delta(r_\mathrm{peri}, r_\mathrm{apo}) = (+3, +47)$~kpc. These shifts are due to the fact that more massive Milky Way-like galaxies have higher central densities, leading to satellites being stripped at larger distances. In other words, the tidal radii of the satellites are smaller, and they are stripped when their density is only marginally higher than that of the host galaxy. As shown in Appendix Fig.~\ref{fig:pa_grid_app_radiusscale}, the results between the low-mass and high-mass groups are consistent with each other when the pericenters and apocenters of the debris structures are rescaled by the virial radius of the host.  

\subsubsection{Puffier Disk}\label{sec:puffier_disk}
Studies of tidal stripping of satellites have found that the presence of a disk in the host galaxy can cause significant disruption of infalling satellite galaxies~\citep[e.g.][]{garrisonkimmel2017}. It follows that the abundance and orbital properties of tidal debris may be affected by the height of the the host galaxy's disk. As such, we consider how these results change when making the disk ``puffier'' than that in the fiducial model.  Specifically, we decrease 
the ratio of the disk scale radius to disk scale height from 25 to 11~(or a factor of $\roughly 2.3$) without changing the amount of mass in the disk.  This amounts to scaling up the disk height. The results are shown by the yellow bars and crosses in Figs.~\ref{fig:numbers_grid} and \ref{fig:peri_apo_grid}, respectively.

The number of intacts, stream-like debris, and shell-like debris is not distinguishable between the fiducial and puffier-disk models---at least within the large halo-to-halo variance that is observed.  That being said, there are small shifts between the median numbers of intacts and stream-like debris, suggesting that satellites remain intact and lose less mass in the puffier-disk model. As satellites approach the disk, the average density of the host, $\Bar{\rho}(r)$, is smaller than in the fiducial model, causing the dynamical time (Eq.~\ref{eq:dyn_time}) to be larger. The mass loss accordingly decreases. Additionally, if the satellite is within the puffier disk, the tidal radius increases, so mass loss accordingly decreases. 

The shift in median $r_\mathrm{peri}$ and $r_\mathrm{apo}$ for intact satellites between the fiducial and puffier-disk models is largely negligible for lower-mass hosts; for higher-mass hosts,~$\Delta(r_\mathrm{peri}, r_\mathrm{apo}) = (-11,-26) ~\mathrm{kpc}$. Stream-like debris  shift down by a few kpc in $r_\mathrm{peri}$ and $r_\mathrm{apo}$ in both high-mass and low-mass hosts, as compared to the fiducial model. Because the disk is less concentrated in the puffier-disk model,  satellites can survive closer pericentric passages and are less disrupted. This leads to slightly lower $r_\mathrm{peri}$ and $r_\mathrm{apo}$ values as compared to the fiducial  debris. 

\subsubsection{More Massive Disk}\label{sec:heavy_disk}
We now consider the consequences of increasing the mass of the host's stellar disk. In the \SatGen\xspace model, the total potential of the host galaxy is given by the sum of thes potential, with mass $M_\mathrm{vir}$, plus the MN disk potential, with a mass of $M_d = f_d \times M_\mathrm{vir}$, where $f_d$ is the disk fraction. In the fiducial model, $f_d = 0.05$, and in the more-massive-disk model, $f_d = 0.1$. This means that the total mass of the host galaxy increases, but its virial radius stays the same as the fiducial model; the scale length and height of the disk also remain the same as the fiducial model.  The results of this variation are shown by the green bars and crosses in Figs.~\ref{fig:numbers_grid} and \ref{fig:peri_apo_grid}, respectively.

Fig.~\ref{fig:numbers_grid} shows that the fiducial and more-massive-disk models are indistinguishable within halo-to-halo variance.  However, looking at the median number of counts for each system, the more-massive-disk model leads to more stream-like debris compared to the number of intacts.  This suggests that satellites may be  more likely to disrupt for this model. How a host galaxy's disk affects the survivability of its subhalo population has been investigated by \cite{garrisonkimmel2017} using the FIRE simulations.  They simulated a galaxy with twice the mass in the disk and found fewer subhalos, especially at $r \lesssim 70$~kpc, concluding that subhalo depletion most directly correlates with the mass of the central disk. Their results agree with the trends observed in our  suites.  As shown in Fig.~\ref{fig:peri_apo_grid}, intact satellites and stream-like debris shift up by a few-$\mathcal{O}(10)$~kpc in $r_\mathrm{peri}$ and $r_\mathrm{apo}$ in both high-mass and low-mass host galaxies, when comparing the heavier-mass disk to the fiducial model. The trend towards larger $r_\mathrm{peri}$ and $r_\mathrm{apo}$ arises because the more-massive disk allows satellites to be disrupted farther out in the halo due to the decreased dynamical time (meaning that these satellites lose more mass farther out in the halo) combined with the smaller tidal radius of the satellite as it orbits.

\subsubsection{Smoother Feedback}\label{sec:apostle}

Studies of tidal stripping of satellites have found that their inner density profiles---i.e. the presence of dense cusps or low-density cores---can significantly affect their tidal evolution \citep[e.g.][]{Penarrubia_2010, Errani_2022b}.  It follows that the abundance and orbital properties of tidal debris may be affected by the degree to which satellites are cusped or cored.  The main driver of coring in dwarf satellites is believed to come from stellar feedback \citep[e.g.][]{Pontzen_2012}.  As such, we modify the feedback prescription from the NIHAO-like model~\citep{10.1093/mnras/stv2856,10.1093/mnras/staa2790} to the smooth feedback of the APOSTLE simulations~\citep{Sawala_2016}. The results of this variation are shown by the red bars and crosses in Figs.~\ref{fig:numbers_grid} and \ref{fig:peri_apo_grid}, respectively. 

In \SatGen, the feedback model is a function of $M_*/M_\mathrm{vir}$ and is therefore sensitive to the choice of SMHM relation. The effect of feedback on subhalos is implemented as satellite properties are initialized at infall, which consequently affects the satellite's tidal evolution after infall as well. The central density of satellites are reduced by lowering the concentration and the slope of the density profile at small radii \citep[see \S2.3.3 of][]{Jiang_2021}. The slope of the inner density profile in \texttt{SatGen} is $\roughly 0.8$~(cored profiles) in the fiducial model and $\roughly 1.3$~(cuspy profiles) in the smoother-feedback model for disrupting satellites.

Figure~\ref{fig:numbers_grid} shows that the median abundances of intact debris, stream-like debris, and shell-like debris are consistent between the fiducial model and that with smoother feedback, especially considering the significant halo-to-halo variance. We do find, however, that the smoother-feedback model allows for more satellite orbits to reach pericenter $<4$~kpc, likely because the satellites are more cuspy.

As shown in Fig.~\ref{fig:peri_apo_grid}, the shift in median $r_\mathrm{peri}$ and $r_\mathrm{apo}$ between the smooth-feedback and fiducial model is largely negligible for intacts in lower-mass hosts; for higher-mass hosts,~$\Delta(r_\mathrm{peri}, r_\mathrm{apo}) = (-10,-19) ~\mathrm{kpc}$. In the smoother-feedback model, satellites have cuspier profiles, so it follows that they would be able to fall farther into the host halo before being disrupted \citep{Errani_2021}. This shifts the distribution of intact satellites to lower $r_\mathrm{peri}$ and $r_\mathrm{apo}$. These shifts are likely more prominent for higher-mass hosts because feedback effects are maximal for bright dwarf galaxies~($M_* \sim 10^{7-9}~M_\odot$), and higher-host-mass galaxies are more likely to have satellite galaxies in this mass range. The host galaxies in our sample with  $M_\mathrm{vir} > 10^{12}~M_\odot$ have $\roughly 6\%$ more bright dwarfs than those on the low end of the range. 

The shift in median $r_\mathrm{peri}$ and $r_\mathrm{apo}$ for stream-like debris is negligible. This suggests that the \texttt{StreamGen} morphology metric is not sensitive to differences in a halo's inner slope. As such, the debris classification is only affected by feedback modeling when it impacts satellite counts. Because the smoother-feedback model only affects the number counts of satellites with pericenters $< 4$~kpc (as discussed above), which are explicitly excluded from our debris classification, the stream-like and shell-like debris samples are largely insensitive to changes in feedback modeling.

\section{Discussion and Conclusion}\label{sec:conclusion}

In this paper, we introduce \texttt{StreamGen}, a tool for easily estimating the properties of tidal debris populations around Milky Way-mass hosts. \texttt{StreamGen} is built to analyze satellite galaxies produced by the semi-analytic satellite galaxy generator, \texttt{SatGen}, but in principle is applicable to any host-satellite pair given orbital quantities and is available on \href{https://github.com/adropulic/StreamGen.git}{\texttt{GitHub}}. 

We generate $\roughly 1500$ galaxies in \SatGen\xspace with varying host parameters in the range of current Milky Way estimates---including mass, disk mass fraction, and ratio of disk scale height to scale length---as well as varying feedback model. We classify the satellite galaxies' debris morphology using \texttt{StreamGen}, which is based on a theoretical model of phase mixing from~\cite{hendel_and_johnston}. We compare the abundances and orbital distributions of these tidal debris populations across morphological classes within a single host galaxy, across galaxies in a single model, and across thousands of galaxies across models. We find that:
\begin{itemize}
    \item Halo-to-halo variance dominates the variation in abundance and orbital distribution of disrupting satellites  compared to changes in the host galaxy model.
    \item The abundance of tidal debris is consistent within the halo-to-halo variance across all models. However, the changes in the ratio of stream-like debris to intact satellites is indicative of different disruption rates across models~(i.e. more on average in the more-massive-disk model and less in the puffier-disk model). 
    \item There are differences in the \emph{median} orbital distributions of stream-like debris across models. Median $r_\mathrm{peri}$ and $r_\mathrm{apo}$ of stream-like debris increases with host galaxy mass. Disk properties also affect the median $r_\mathrm{peri}$ and $r_\mathrm{apo}$ of stream-like debris, with more massive~(puffier) disks leading to higher~(lower) median stream $r_\mathrm{peri}$ and $r_\mathrm{apo}$. 
\end{itemize}
These results strongly indicate that the effect of halo-to-halo variance dominates that of variations to the host galaxy's potential within Milky Way estimates on the abundance and orbital properties of tidal debris. It is possible that more significant changes to the host properties---beyond those considered in this work---could dominate halo-to-halo variance within a single model.

As currently defined, the morphology metric is not sensitive to differences in the inner regions of the satellite~($\roughly 0.01$ times the satellite's virial radius), as could be caused by stellar feedback. It is possible that \texttt{StreamGen} could be more sensitive to differences in disruption due to cored vs. cuspy satellite profiles if we considered changes in the morphology metric throughout each satellite's entire lifetime, instead of calculating a single value at p300. These effects, however, should not affect the average morphology of the tidal debris that we are modeling in this work (which is most analogous to the morphology that would be assigned based on observations), only the variation across multiple stripping events. This study of the effect of feedback on semi-analytic stream populations may be complemented by future studies of stream disruption in cosmological and idealized N-body simulations, which would account for the continuous affect of stellar feedback across the lifetime and full disruption history of the satellite.

After incorporating criteria for a stream or shell to be considered detectable, \texttt{StreamGen} could be used to predict the expected spread on the abundances and orbital distributions of observed populations of tidal debris. Given the large samples of host galaxies that \texttt{StreamGen} can generate, it can also be utilized to enhance the interpretability of generative AI models by providing large population models of stream-like and shell-like debris, offering valuable insights into how these structures evolve under various host galaxy conditions. 

The halo-to-halo variance, and to a lesser degree the variations in host properties, observed in our \texttt{StreamGen} suites may explain the discrepancy that~\cite{Shipp_2023} found: FIRE galaxies in the Milky Way-mass range~(${M_{\mathrm{vir}} \sim 10^{12.0-12.25}~M_\odot}$) are ``missing" stream-like debris on orbits at low pericenter ($\lesssim 20$~kpc) and apocenter ($\lesssim 50$~kpc), as compared to the Milky Way. We find streams at lower pericenters and apocenters in all models, but also that certain models can favor lower median pericenters and apocenters of stream populations (e.g., puffier-disk model). 

It is important to keep in mind that \texttt{SatGen} makes a number of assumptions in its semi-analytical approach to satellite modeling. It assumes spherical symmetry for the host and satellites and does not model certain processes such as: disk growth~(except as a function of the total virial mass of the host) or dynamics, the back reaction of the satellites on the host halo and the interaction of satellites with each other. Additionally, the SMHM relation is set from data and the baryonic feedback model is tuned to simulations. Any of these choices can affect the tidal disruption of debris and must be carefully considered before performing direct comparisons to data. The \texttt{StreamGen} pipeline can be further refined by incorporating a time-evolving potential component for an infalling LMC analog, which would contribute to the disruption of satellites~\citep{Erkal:2019, Shipp_2021} and change the velocity distribution of infalling satellites~\citep{arora2023lmc}. Additionally, while globular cluster stream-like debris are not considered in this work, their inclusion presents an avenue for future development in semi-analytic simulations and the study of stream population statistics~\citep{Meng_2022,Chen_2024, pearson2024}. 

Finally, the upcoming Rubin Observatory~\citep{ivezic2019lsst} and the Nancy Grace Roman Space Telescope~\citep{spergel2015wide} are set to significantly enhance the measurement and analysis of faint tidal debris, bracing the field of near-field cosmology for groundbreaking advancements. Of particular interest in the faint tidal debris regime are recent discoveries of extra-tidal stars, or those between 1--5 times the tidal radius of the satellite galaxy or globular cluster, which provide valuable information on the effects of tidal forces~\citep[e.g.][]{Piatti2021,Kundu2022,Xu2024} on satellite populations. Rubin and Roman will reveal further observations of the low-surface brightness outskirts of Milky Way satellites and provide measurements of the disruption rates of satellite populations. 

\texttt{StreamGen} serves as a bridge between these observational data and cosmological simulations by providing a robust framework for exploring the impacts of various galactic parameters and feedback mechanisms on tidal debris. By creating an avenue to rapidly generate and identify populations of tidal debris under different host galaxy conditions, \texttt{StreamGen} can allow observers and simulators alike to gain a deeper understanding of the variance of tidal debris in galaxies, preventing over-interpreted explanations for varying distributions of small-scale structure. Analyzing populations of simulated disrupting satellite galaxies is a critical step in achieving a comprehensive insight into the complexities of galaxy formation and evolution, as well as the fundamental nature of dark matter.

\section*{Acknowledgements} \label{sec:acknowledge}
The authors would like to thank F.~Jiang, D.~Folsom, D.~Erkal, S.~Roy, T.~Nguyen, and R.~Errani for fruitful conversations. AD is supported by NSF~GRFP under Grant No.~DGE-2039656. NS is supported by an NSF Astronomy and Astrophysics Postdoctoral Fellowship under award AST-2303841. LN is supported by the Sloan Fellowship, the NSF CAREER award 2337864, NSF award 2307788, and by the National Science Foundation under Cooperative Agreement PHY-2019786 (The NSF AI Institute for Artificial Intelligence and Fundamental Interactions, http://iaifi.org/). ML and AD are supported by the Department of Energy~(DOE) under Award Number DE-SC0007968.  ML is also supported by the Simons Investigator in Physics Award. This project was developed in part at the Streams24 meeting hosted at Durham University. The simulations presented in this article were primarily performed on computational resources managed and supported by Princeton Research Computing, a consortium of groups including the Princeton Institute for Computational Science and Engineering (PICSciE) and the Office of Information Technology's High Performance Computing Center and Visualization Laboratory at Princeton University. The authors acknowledge the MIT Office of Research Computing and Data for providing high performance computing resources that have contributed to the research results reported within this paper. It made use of the \texttt{astropy}~\citep{Robitaille:2013mpa},
\texttt{Jupyter}~\citep{Kluyver2016JupyterN}, \texttt{matplotlib}~\citep{Hunter:2007}, 
\texttt{NumPy}~\citep{numpy:2011}, \texttt{pandas}~\citep{mckinney-proc-scipy-2010}, \texttt{SciPy}~\citep{Jones:2001ab}, and \texttt{agama}~\cite{vasiliev2019agama} software packages.
\section*{Data Availability}
\SatGen\xspace is publicly available on \href{https://github.com/shergreen/SatGen.git}{\texttt{GitHub}}. \texttt{StreamGen} is publicly available on \href{https://github.com/adropulic/StreamGen.git}{\texttt{GitHub}}.

\clearpage

\bibliography{satgen_streams}{}
\bibliographystyle{aasjournal}

\clearpage

\appendix

\setcounter{equation}{0}
\setcounter{figure}{0} 
\setcounter{table}{0}
\renewcommand{\theequation}{A\arabic{equation}}
\renewcommand{\thefigure}{A\arabic{figure}}
\renewcommand{\thetable}{A\arabic{table}}
\renewcommand*{\theHfigure}{\thefigure}
\renewcommand*{\theHtable}{\thetable}
\renewcommand*{\theHequation}{\theequation}

\section{Morphology Metric Uncertainty Analysis}
\label{app:uncertainty}

We perform a series of tests to assess the impact of any uncertainty in the morphology assignment on our conclusions~(i.e. if and how the distributions would change due to the uncertainty in the morphology metric prediction). In Sec.~\ref{sec:validate_metric}, we validated the morphology metric by performing N-body simulations, visually inspecting the resulting debris, and comparing it to the morphology metric assignment. Similarly to HJ15, we found ``uncertain" debris, or that which is assigned a stream-like or shell-like morphology by the metric even though the visual categorization of the debris is nebulous. In Fig.~\ref{fig:validation} for example, this is shown by the brown star in the left-hand plot at $\Psi^1_E \sim 15$ deg, which does not look particularly shell-like or stream-like but is too centrally dense to be completely phase mixed.

\begin{figure*}[ht]
    \centering
    \includegraphics[width=0.31\textwidth]{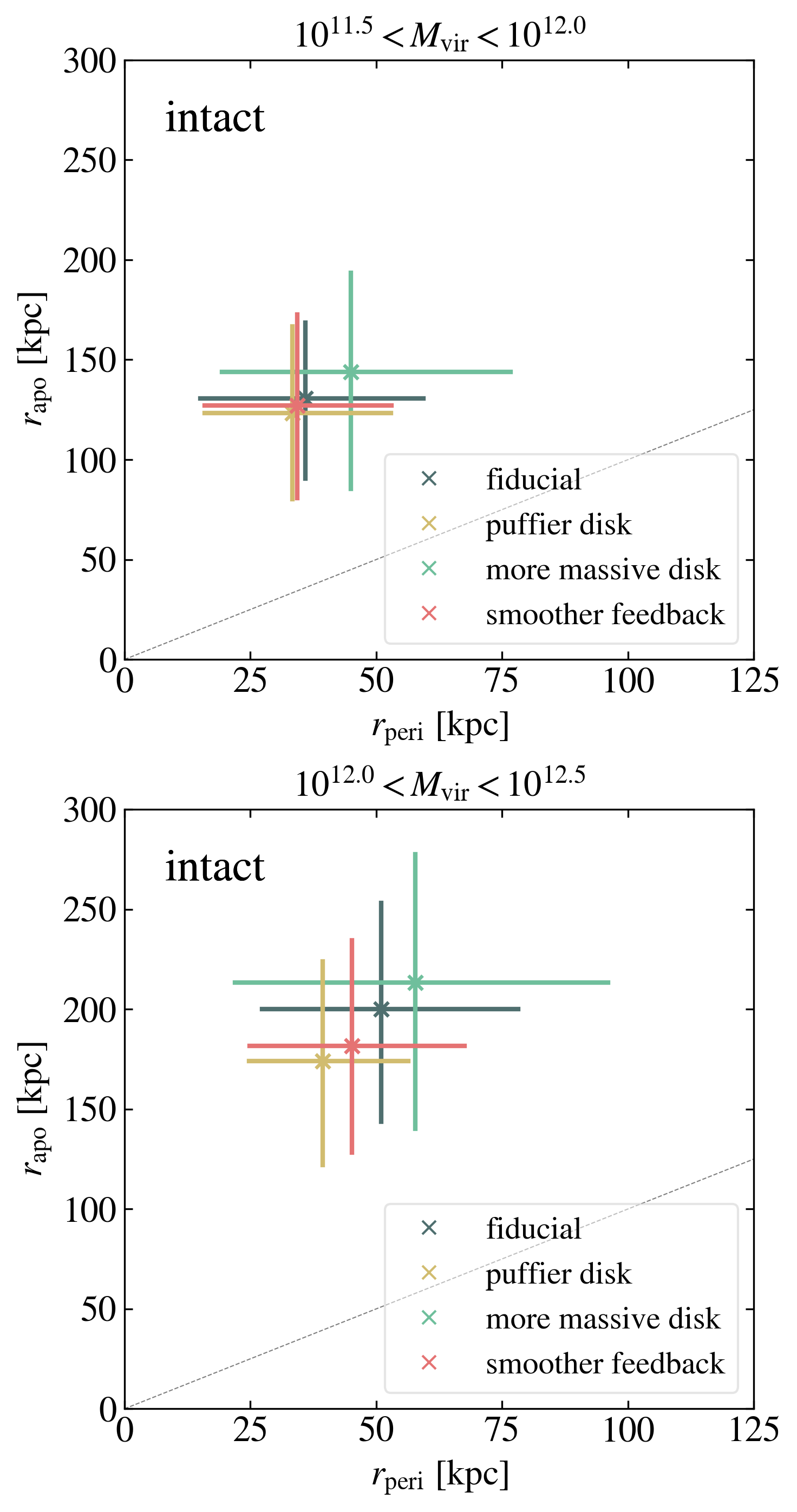}
    \includegraphics[width=0.31\textwidth]{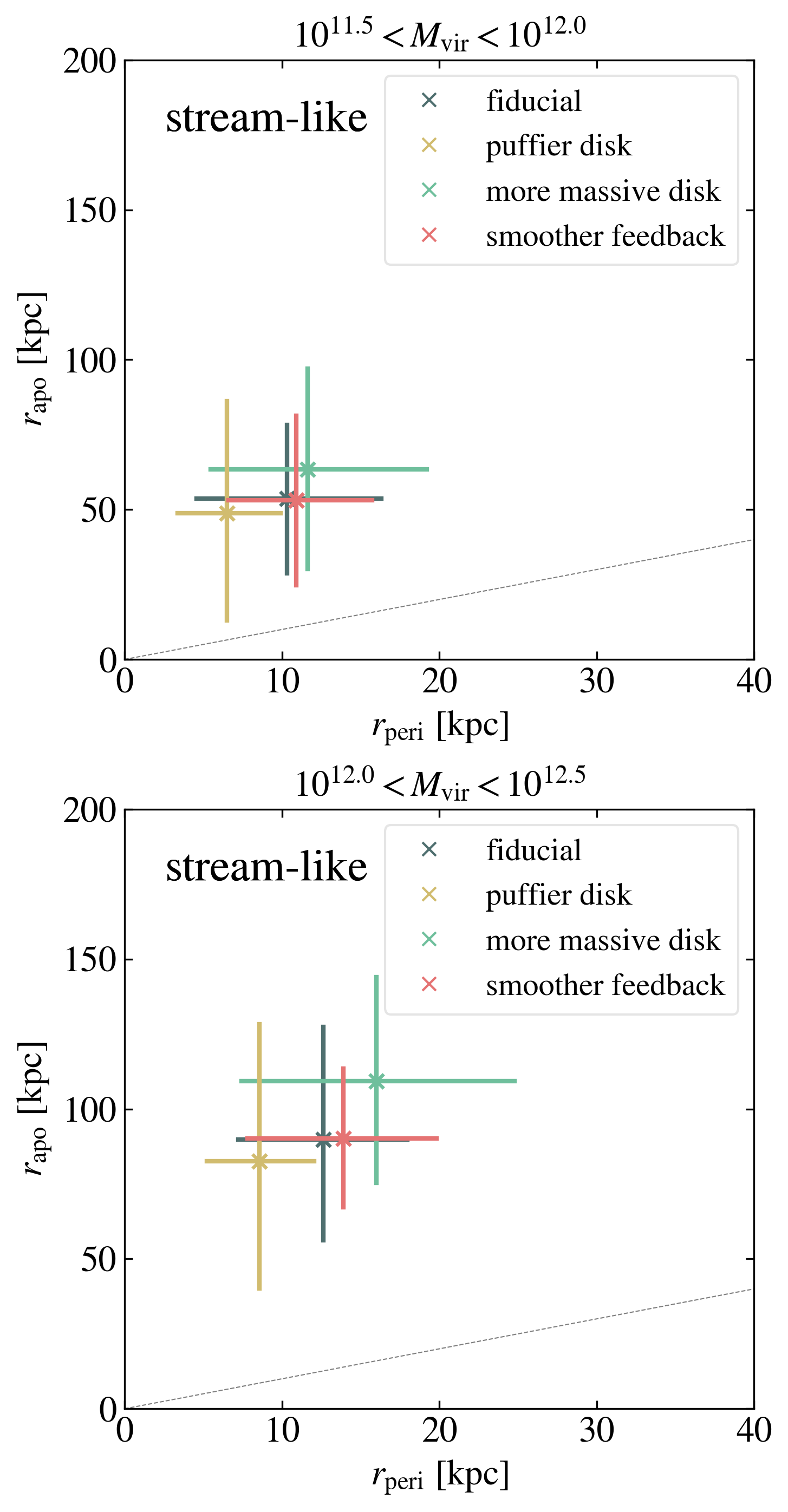}
    \includegraphics[width=0.31\textwidth]{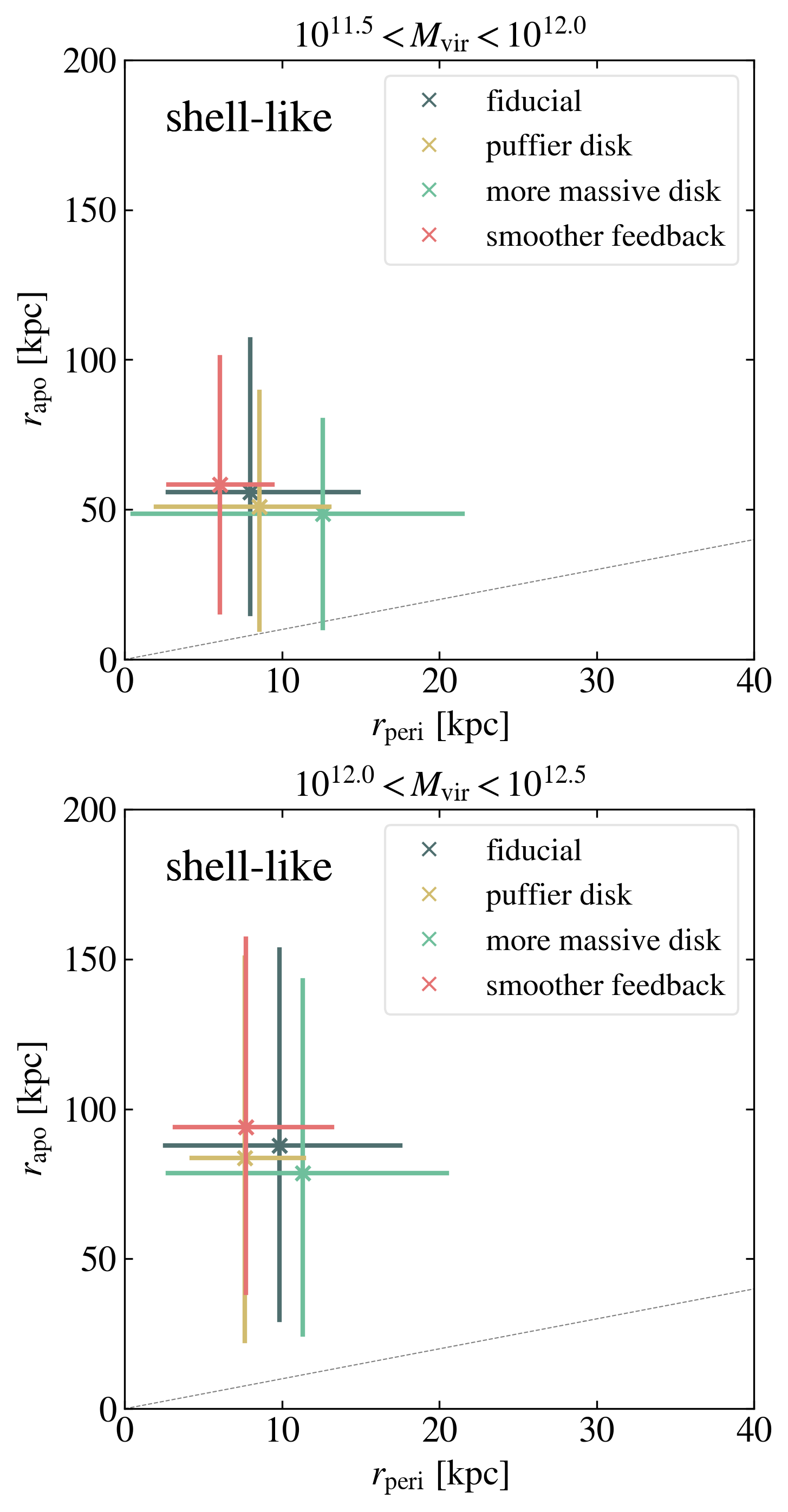} 
    \caption{The pericenter-apocenter distributions for intact satellites~(first column), stream-like debris~(second column), and shell-like debris~(third column), weighted by the procedure described in Appendix~\ref{app:uncertainty}. This plot is otherwise identical to Fig.~\ref{fig:peri_apo_grid}. Accounting for the uncertainty in classifying the tidal debris, as specified by HJ15, does not change the conclusions.}
    \label{fig:pa_grid_app}
\end{figure*}

To determine whether these ``uncertain'' classifications impact our final results, we  introduce weights into the analysis. Using HJ15's Fig.~7, which shows the morphology in $\Psi^1_E$ versus $\Psi_L$ space for a large number of tidal debris classified in their simulations as ``stream,'' ``shell,'' or ``unsure.'' We divide the plot in HJ15 into a $9 \times 9$ evenly-spaced grid in the range of $\Psi^1_E < 120$~deg and $\Psi_L < 120$~deg. We then calculate the fraction of uncertain to known debris in each bin.  These bin values are used as the weights for each satellite in our analysis. All points with $\Psi^1_E > 120 $~deg and $\Psi_L > 120$~deg receive a weight of 1.0. The purpose of this procedure is to down-weight the contribution of any points in the bottom-left corner of the $\Psi^1_E$ versus $\Psi_L$ plot, whose classification is most likely to be uncertain. The lowest weight given is 0.6. Comparing Fig.~\ref{fig:pa_grid_app} to  Fig.~\ref{fig:peri_apo_grid}, we find that the pericenter and apocenter distributions do not change significantly when accounting for this uncertainty.

\clearpage
\section{Additional Plots}
\label{app:additional_plots}

\setcounter{equation}{0}
\setcounter{figure}{0} 
\setcounter{table}{0}
\renewcommand{\theequation}{B\arabic{equation}}
\renewcommand{\thefigure}{B\arabic{figure}}
\renewcommand{\thetable}{B\arabic{table}}
\renewcommand*{\theHfigure}{\thefigure}
\renewcommand*{\theHtable}{\thetable}
\renewcommand*{\theHequation}{\theequation}

\begin{figure*}[ht]
    \centering
    \includegraphics[width=0.30\textwidth]{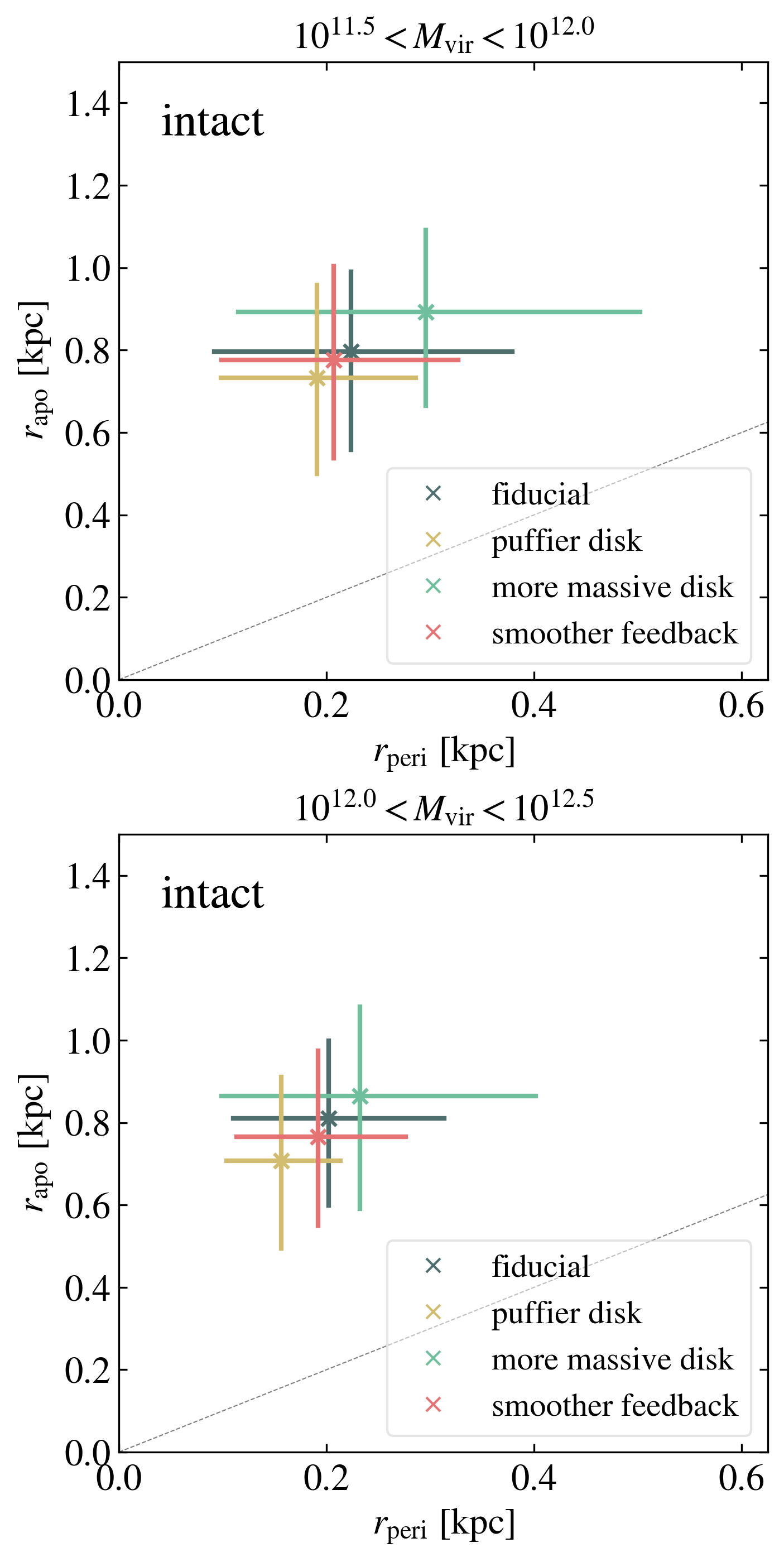}
    \includegraphics[width=0.31\textwidth]{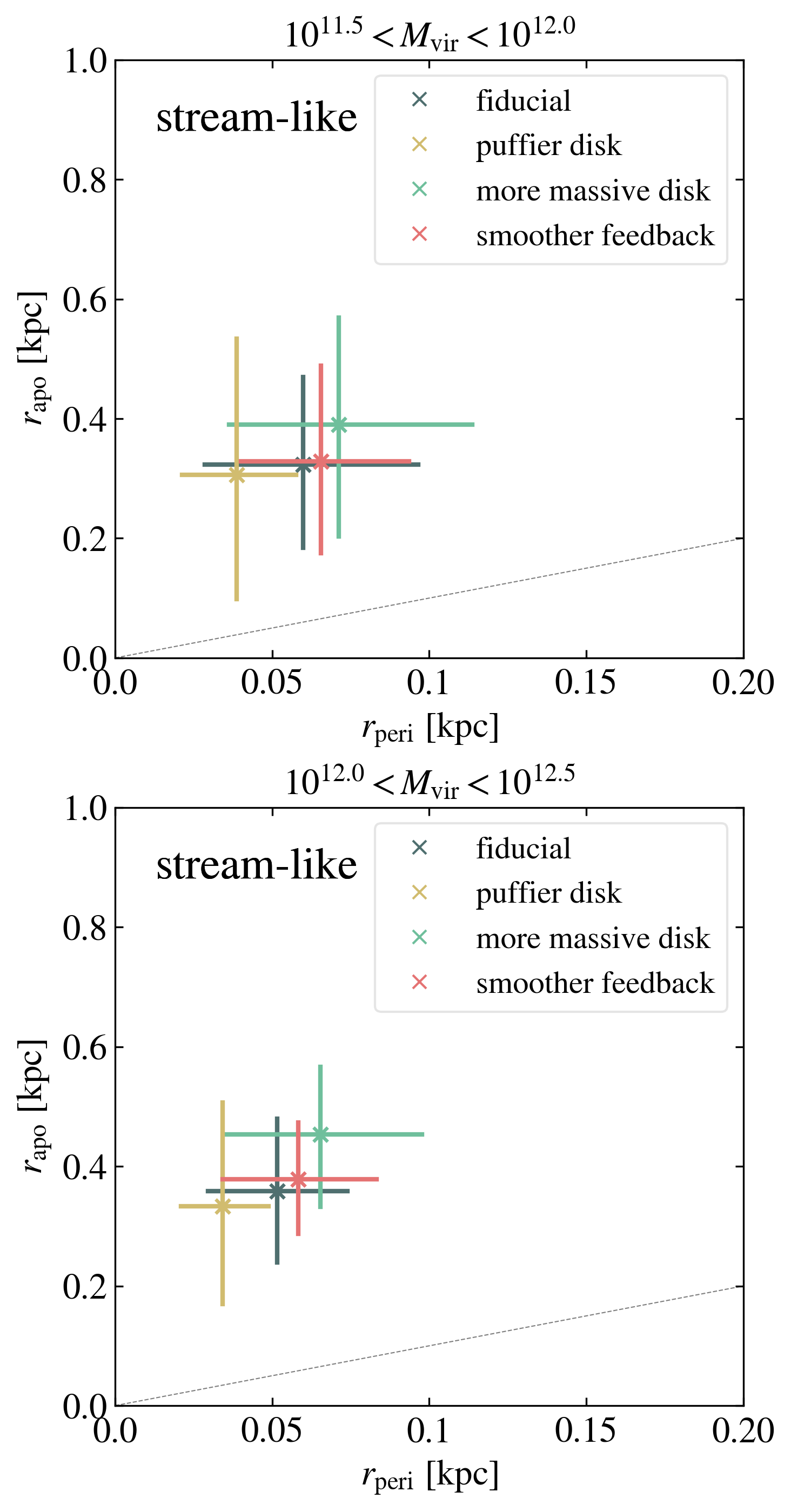}
    \includegraphics[width=0.31\textwidth]{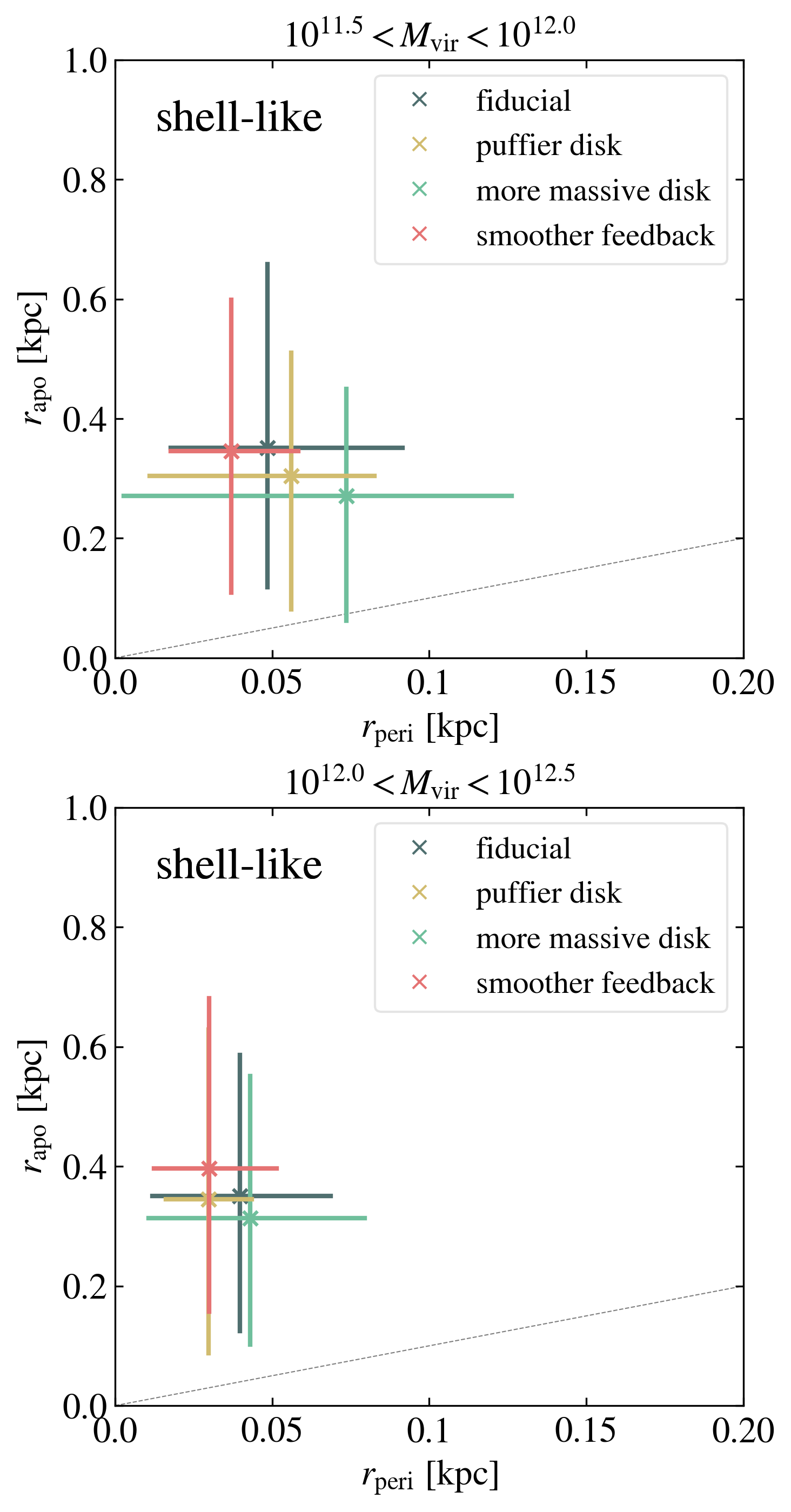} 
    \caption{The pericenter-apocenter distributions for intact satellites~(first column), stream-like debris~(second column), and shell-like debris~(third column). This plot is identical to Fig.~\ref{fig:peri_apo_grid}, except that the pericenters and apocenters have been scaled by the virial radius of the host. With this scaling, there is no significant offset in the median pericenters and apocenters of each debris type between the low- and high-mass host ranges.}
    \label{fig:pa_grid_app_radiusscale}
\end{figure*}

\end{document}